\documentclass[twocolumn, trackchanges]{aastex631}

\usepackage{graphicx}
\usepackage{multirow}
\usepackage{appendix}

\begin{document}

\title{Radio Activity Across Accretion State Changes in Changing-look AGNs: Insights from FIRST and VLASS over Two Decades}

    \author[0009-0008-3338-393X]{Zhi-Qiang Chen} 
    \affiliation{School of Physics and Technology, Nanjing Normal University, No. 1,
	Wenyuan Road, Nanjing, 210023, P. R. China\\ Email:\href{mailto:yuanqirong@njnu.edu.cn}{yuanqirong@njnu.edu.cn}}

    \author[0000-0001-9457-0589]{Wei-Jian Guo}
    \affiliation{Key Laboratory of Optical Astronomy, National Astronomical Observatories, Chinese Academy of Sciences, Beijing 100012, P.R. China\\ Email:\href{mailto:guowj@bao.ac.cn，zouhu@nao.cas.cn}{guowj@bao.ac.cn, zouhu@nao.cas.cn}}

    \author[0000-0003-1251-532X]{Victoria A. Fawcett}
    \affiliation{European Southern Observatory, Karl-Schwarzschild-Stra\ss e~2, 85748~Garching~bei~M\"unchen, Germany
}

    \author[0000-0002-8402-3722]{Jun-Jie Jin}
    \affiliation{Key Laboratory of Optical Astronomy, National Astronomical Observatories, Chinese Academy of Sciences, Beijing 100012, P.R. China\\ Email:\href{mailto:guowj@bao.ac.cn，zouhu@nao.cas.cn}{guowj@bao.ac.cn, zouhu@nao.cas.cn}}

    \author[0000-0002-3742-6609]{Wen-Ke Ren}
    \affiliation{Shanghai Astronomical Observatory, Chinese Academy of Sciences, 80 Nandan Road, Shanghai 200030, P.R. China}

    \author[0000-0001-8416-7059]{Heng-Xiao Guo}
    \affiliation{Shanghai Astronomical Observatory, Chinese Academy of Sciences, 80 Nandan Road, Shanghai 200030, P.R. China}

    \author[0000-0002-4455-6946]{Min-Feng Gu}
    \affiliation{Shanghai Astronomical Observatory, Chinese Academy of Sciences, 80 Nandan Road, Shanghai 200030, P.R. China}

    \author[0000-0003-3226-031X]{Yan-Mei Chen}
    \affiliation{School of Astronomy and Space Science, Nanjing University, Nanjing 210023, P.R. China}
    \affiliation{Key Laboratory of Modern Astronomy and Astrophysics (Nanjing University), Ministry of Education, Nanjing 210023, P.R. China}

    \author{Lu Feng}
    \affiliation{Key Laboratory of Optical Astronomy, National Astronomical Observatories, Chinese Academy of Sciences, Beijing 100012, P.R. China\\ Email:\href{mailto:guowj@bao.ac.cn，zouhu@nao.cas.cn}{guowj@bao.ac.cn, zouhu@nao.cas.cn}}

    \author[0000-0002-6684-3997]{Hu Zou}
    \affiliation{Key Laboratory of Optical Astronomy, National Astronomical Observatories, Chinese Academy of Sciences, Beijing 100012, P.R. China\\ Email:\href{mailto:guowj@bao.ac.cn，zouhu@nao.cas.cn}{guowj@bao.ac.cn, zouhu@nao.cas.cn}}

    \author[0000-0002-9244-3938]{Qi-Rong Yuan}
    \affiliation{School of Physics and Technology, Nanjing Normal University, No. 1,
	Wenyuan Road, Nanjing, 210023, P. R. China\\ Email:\href{mailto:yuanqirong@njnu.edu.cn}{yuanqirong@njnu.edu.cn}}
    \affiliation{University of Chinese Academy of Sciences, Nanjing, 211135, P. R. China}

\begin{abstract}

Changing-look active galactic nuclei (CL-AGNs) provide a unique opportunity to probe the coupling between accretion flows and relativistic jets in supermassive black holes. We investigate the long-term radio behavior of CL-AGNs over $\sim$20 years by combining FIRST and VLASS observations with quasi-simultaneous optical spectroscopy and photometry. From a parent sample of 1092 CL-AGNs, we identify 58 sources with radio detections. Radio-detected CL-AGNs exhibit systematically higher radio kinetic efficiency, quantified by $P_{\rm j}/L_{\rm bol}$, than both typical radio-detected AGNs and radio transients, consistent with their preference for low Eddington ratios. At the population level, the expected anti-correlation between radio emission and accretion rate is weak. However, a clear source-by-source anti-correlation emerges in a small subset of CL-AGNs with continuous multi-epoch coverage. We further identify four radio transients, including both radio turn-on and turn-off events, and one source exhibiting a multiwavelength flare that may be indicative of tidal disruption event–like activity.
These results suggest that radio activity in CL-AGNs is not governed by instantaneous accretion state changes but is instead regulated by long-term accretion history and jet evolution, with additional stochastic or transient channels contributing in rare cases.

\end{abstract}


\section{INTRODUCTION}  \label{sec:intro}

Supermassive black holes (SMBHs) reside at the centers of most massive galaxies, and when actively accreting, they manifest as active galactic nuclei (AGNs) that radiate across the entire electromagnetic spectrum \citep{1998AJ....115.2285M}. Among the various observational windows, radio emission plays a uniquely important role in probing the physical processes in AGN central engines, as it is largely insensitive to dust obscuration and directly traces non-thermal emission associated with relativistic particles. In most AGNs, the radio emission is dominated by synchrotron radiation produced by relativistic electrons in collimated jets or compact nuclear outflows \citep{1979ApJ...232...34B,1995ApJ...450..559B}. By contrast, radio emission associated with star formation arises primarily from supernova remnants and thermal free--free processes, is spatially extended on kiloparsec scales, and is typically subdominant in the nuclear regions of AGNs \citep{1992ARA&A..30..575C}. As a result, radio observations provide a powerful and relatively clean diagnostic of jet activity and its connection to accretion onto SMBHs.

A growing body of observational and theoretical work suggests that jet production in accreting black hole systems is fundamentally linked to the properties of the accretion flow (e.g., \citealt{2003MNRAS.345.1057M,2007MNRAS.375..513M,2014MNRAS.445...81S,2018A&A...615A..57M}). In AGNs, the strength of radio emission and jet production efficiency have been shown to depend on both black hole mass and accretion rate, often parameterized by the Eddington ratio. In particular, systems accreting at relatively low Eddington ratios tend to exhibit stronger radio emission and more efficient jet production. This behavior is commonly interpreted within a disk--jet coupling framework, in which radiatively inefficient accretion flows are more effective at launching and sustaining relativistic jets, whereas radiatively efficient thin disks tend to suppress large-scale jet activity (e.g., \citealt{2006MNRAS.372.1366K,2007ApJ...658..815S}).

The disk--jet coupling is most clearly established in stellar-mass black hole X-ray binaries (XRBs), where transitions between distinct accretion states can be directly observed on human-accessible timescales. In these systems, the low/hard state is characterized by steady compact jets and strong radio emission, whereas the high/soft state exhibits quenched radio emission and weak or absent jets, with the transition occurring at a characteristic Eddington ratio of $L_{\rm bol}/L_{\rm Edd} \sim 0.01$ \citep{2003A&A...409..697M,2004MNRAS.355.1105F}. Empirical correlations between radio and X-ray luminosities further demonstrate a tight coupling between inflow and outflow processes in XRBs \citep{2003A&A...400.1007C}. These results have motivated the idea that a scale-invariant disk--jet coupling may operate across many orders of magnitude in black hole mass, extending from stellar-mass black holes to SMBHs.

Despite this theoretical expectation, direct observational evidence for XRB-like accretion state transitions and their associated jet responses in AGNs remains elusive. This difficulty arises primarily from two factors. First, most AGNs exhibit relatively stable accretion rates, and changes large enough to induce a transition in the accretion mode are rarely observed \citep{2020MNRAS.492.2335L}. Second, while optical and ultraviolet emission can respond rapidly to changes in accretion rate, radio emission generally evolves on much longer timescales, reflecting the cumulative or time-integrated nature of jet activity. As a result, any jet response to short-term accretion variability may be diluted or smeared out in the radio band, complicating efforts to establish a clear causal connection between accretion state changes and radio emission in AGNs.

Changing-look AGNs (CL-AGNs), which undergo dramatic transitions between Type~1 (showing both broad and narrow emission lines) and Type~2 (showing only narrow emission lines) spectral classifications on timescales of years to decades, provide a unique opportunity to overcome these limitations. These transitions are widely interpreted as being driven by substantial changes in accretion rate rather than variable obscuration (e.g., \citealt{2015ApJ...800..144L,2017ApJ...846L...7S,2023ApJ...953...61Y}). Such changes may be triggered by rapid variations in mass supply, including tidal disruption event (TDE)--like episodes \citep{2017ApJ...843..106B,2022ApJ...933...70L}, or by intrinsic accretion instabilities that alter the structure of the accretion flow \citep{2020MNRAS.492.5540M,2020A&A...641A.167S}. As a result, CL-AGNs are often regarded as the supermassive analogs of XRBs, exhibiting coordinated variability in continuum emission, broad emission lines, and in some cases X-ray properties.

If disk--jet coupling operates in a scale-invariant manner, CL-AGNs should provide a natural laboratory for testing whether relativistic jets in AGNs respond coherently to rapid changes in accretion state. However, whether changing-look transitions directly trigger, suppress, or otherwise regulate jet activity remains an open question. In particular, the long response timescale of radio emission raises the possibility that jet activity reflects long-term averaged accretion properties rather than instantaneous accretion states. Addressing this issue requires long-term radio observations combined with contemporaneous constraints on accretion properties.

In this work, we investigate the radio behavior of CL-AGNs over a timescale of approximately 20 years by combining observations from the the Faint Images of the Radio Sky at Twenty-cm (FIRST) and the Karl G. Jansky Very Large Array Sky Survey (VLASS) radio surveys with quasi-simultaneous optical spectroscopy and photometry. By comparing CL-AGNs with typical AGNs and radio transients, and by examining both population-level statistics and source-by-source evolutionary behavior, we aim to assess the nature of disk--jet coupling in CL-AGNs and to determine whether changing-look transitions are directly linked to the launching or cessation of relativistic jets.

The paper is structured as follows. In Section~\ref{sec:Data}, we describe the radio catalogs used in this work and the sample selection, including the CL-AGN sample, the typical AGN control sample, and the radio transient sample. Section~\ref{Sec: method and result} presents the spectral fitting procedure, the method for deriving the monochromatic luminosity $L_{5100}$ from optical photometric data, and the statistical radio properties of the CL-AGNs. A comprehensive analysis of the radio properties of CL-AGNs is carried out in Section~\ref{Sec: discussion}, followed by a brief summary and conclusions in Section~\ref{Sec: Summary}. Throughout this study, a concordance $\Lambda$CDM cosmology is adopted, with parameters $\Omega_{m}=0.32$, $\Omega_{\Lambda}=0.68$, and $H_{0}=67~{\rm km~s^{-1}~Mpc^{-1}}$ \citep{2020A&A...641A...6P}.

\section{Data and Sample}\label{sec:Data}

In this section, we describe the radio data sets and sample construction adopted in this work. Our analysis is based on two wide-area radio surveys, FIRST and VLASS, which provide a uniform time baseline of approximately two decades for probing long-term radio evolution in active galactic nuclei. Using this radio framework, we construct a primary sample of CL-AGNs by cross-matching a comprehensive literature compilation with the radio catalogs. To place the radio properties of CL-AGNs in context, we further define two comparison samples: (i) a control sample of typical radio-detected Type~1 AGNs, representing baseline jet behavior in the absence of changing-look transitions, and (ii) a sample of radio transients, tracing extreme cases of rapid jet emergence or fading. This design enables us to compare population-level trends and source-by-source evolution, and to assess whether the radio properties of CL-AGNs are distinct from those of normal AGNs and radio transient populations.

\subsection{Radio Surveys}

Below we summarize the basic characteristics of the FIRST and VLASS surveys, including their frequency coverage, angular resolution, sensitivity, and catalog products used in this work.
The comparable sensitivities and complementary epochs of FIRST and VLASS allow us to trace radio emission from AGNs over decade-long timescales, providing a suitable framework for investigating long-term jet evolution.

FIRST is a systematic radio survey conducted at 1.4~GHz using the Very Large Array (VLA) in its B configuration \citep{1994ASPC...61..165B}. It covers a total sky area of 10,575~deg$^{2}$ in the North and South Galactic Caps, with observations carried out between 1993--2004 and 2009--2011 \citep{1995ApJ...450..559B,2015ApJ...801...26H}. The survey has an angular resolution of approximately $5^{\prime\prime}$ and a typical rms sensitivity of 0.15~mJy~beam$^{-1}$. The FIRST source catalog has a detection threshold of $\sim$1~mJy and provides both peak and integrated flux densities obtained from two-dimensional Gaussian fitting. FIRST image cutouts and catalogs are publicly available through the FIRST data archive.\footnote{\url{http://sundog.stsci.edu/}}

VLASS is a more recent wide-area radio survey carried out at S-band (2--4~GHz) with the VLA, designed to map approximately 80\% of the sky at declinations $\gtrsim -40^{\circ}$ over three epochs separated by $\sim$32 months \citep{2020PASP..132c5001L}. In this work, we use data from the first two epochs, with epoch~1 spanning September~2017 to July~2019 and epoch~2 spanning June~2020 to March~2022. VLASS covers a total area of 33,885~deg$^{2}$ with an angular resolution of $\sim2.5^{\prime\prime}$ and a per-epoch rms sensitivity of $\sim$0.12~mJy~beam$^{-1}$, comparable to FIRST. Source detection and flux measurements are performed on the Quick Look images using the Python Blob Detector and Source Finder (\textsc{PyBDSF}; \citealt{2015ascl.soft02007M}), as described by \citet{2021ApJS..255...30G}. The VLASS catalogs and image products are publicly available via the CIRADA and NRAO archives.\footnote{\url{https://cirada.ca/catalogs}}\footnote{\url{https://archive-new.nrao.edu/vlass/quicklook/}}

LoTSS is a wide-area low-frequency radio survey conducted with the LOw Frequency ARray (LOFAR) at 120--168~MHz (central frequency $\simeq144$~MHz), providing sensitive continuum imaging over large regions of the northern sky \citep{2022A&A...659A...1S}. The LoTSS second data release (DR2) reaches an angular resolution of $\sim6^{\prime\prime}$ and a typical rms sensitivity of $\sim(70$--$100)\,\mu$Jy~beam$^{-1}$, enabling reliable measurements of integrated 144~MHz flux densities for faint radio AGN \citep{2022A&A...659A...1S}. Source extraction in the LoTSS catalogs is performed on the LoTSS images and provides integrated flux densities and associated uncertainties for detected sources. LoTSS image products and catalogs are publicly available through the LOFAR surveys archive.\footnote{\url{https://lofar-surveys.org/}}
In this work, LoTSS provides complementary low-frequency constraints that are particularly useful for characterizing radio spectral shapes. For radio-detected CL-AGNs with LoTSS counterparts, we combine the 144~MHz flux densities with the VLASS 3~GHz measurements to estimate broadband radio spectral indices (Section~\ref{Sec: radio transients}), while noting that the non-simultaneity of LoTSS and VLASS observations may introduce additional scatter due to intrinsic variability.

\subsection{Changing-look AGN Sample}

We construct a parent sample of changing-look AGNs by compiling all reported CL-AGNs from the literature. This compilation includes 1092 sources identified through optical spectroscopic variability and multi-wavelength diagnostics (e.g., \citealt{2015ApJ...800..144L,2016MNRAS.457..389M,2016ApJ...826..188R,2018ApJ...862..109Y,2020ApJ...889...46S,2022ApJ...933..180G,2023MNRAS.524..188L,2024ApJ...966...85Z,2024ApJS..270...26G,2025ApJS..278...28G,2025ApJ...980...91Y,2025ApJ...986..160D,2026ApJS..282...28C}). This parent sample represents the largest available collection of confirmed CL-AGNs.

We cross-match the parent CL-AGN sample with the FIRST and VLASS catalogs using matching radii of $5^{\prime\prime}$ and $3^{\prime\prime}$, respectively, reflecting the different angular resolutions of the two surveys. To minimize contamination from spurious detections and imaging artifacts, we require the matched radio sources to have flux densities exceeding 2~mJy. Applying these criteria, we identify 58 CL-AGNs with radio detections spanning the $\sim$20-year baseline defined by FIRST and VLASS. These sources constitute our primary working sample and are listed in Table~\ref{tab:radio_clagn}.

Among the radio-detected CL-AGNs, 54 sources are detected in both FIRST and at least one epoch of VLASS, indicating persistent radio emission over decades. The remaining four sources exhibit the sudden appearance or disappearance of radio emission between FIRST and VLASS, corresponding to dramatic radio variability. Following previous studies, we classify these objects as radio transients \citep{2020ApJ...905...74N,2021ApJ...914...22W,2022ApJ...938...43Z}. This small subset allows us to explore extreme cases of jet emergence or cessation within the CL-AGN population.

Given the potential connection between radio activity and X-ray emission from the innermost accretion regions, we also examined the availability of archival X-ray observations for the radio-detected CL-AGN sample.
To assess the availability of X-ray observations, we cross-matched the 58 radio-detected CL-AGNs with the Second ROSAT All-Sky Survey Point Source Catalog (2RXS), the XMM-Newton DR14 source catalog, and the Chandra Source Catalog Release 2.1 using a matching radius of 10\arcsec. We identify 8, 6, and 3 counterparts in the ROSAT, XMM-Newton, and Chandra catalogs, respectively. After accounting for duplicate matches between catalogs, a total of 15 unique CL-AGNs possess X-ray detections in at least one X-ray survey, corresponding to $\sim26\%$ of the radio-detected sample.
The relatively small fraction of X-ray detections is likely attributable, at least in part, to the incomplete sky coverage of current X-ray surveys, particularly the pointed observations conducted by XMM-Newton and Chandra, rather than implying an intrinsic lack of X-ray emission from the CL-AGN population. We therefore do not attempt a detailed investigation of the X-ray properties of these sources in this work. A comprehensive X-ray characteristics analysis of the CL-AGNs will be presented in a future study.

\subsection{Control Samples}

To assess whether the radio properties of CL-AGNs are distinct from those of other AGN populations and to isolate the physical drivers of their radio emission, we construct three comparison samples: a control sample of radio silent CL-AGNs, a control sample of typical AGNs, and a sample of radio transients.

\subsubsection{Radio Silent CL-AGNs}\label{Sec: radio silent sample}

To understand the fundamental triggering conditions of radio activity, it is crucial to isolate the physical mechanisms driving the radio emission from the intrinsic changing-look phenomenon itself. For this purpose, we construct a baseline control sample of 536 radio-silent CL-AGNs. These sources are directly drawn from the recent CL-AGN catalog compiled by \cite{2025ApJS..278...28G}, specifically selecting those that lack any radio detections in both the FIRST and VLASS surveys. By comparing this radio-silent population with our radio-detected sample, we can systematically investigate whether fundamental parameters, such as black hole mass, accretion rate, and host-galaxy star formation activity, inherently differ between the radio-active and radio-quiet states of CL-AGNs.

\subsubsection{Typical AGN Sample}

We construct a control sample of typical AGNs to represent baseline jet behavior in the absence of changing-look transitions. Type~1 AGNs are selected from the Sloan Digital Sky Survey (SDSS) DR14 quasar catalog \citep{2020ApJS..249...17R}. For sources with redshift $z<0.6$, we require the full width at half maximum (FWHM) of the H$\beta$ emission line to be $\geq 2500~\mathrm{km~s^{-1}}$, while for sources at $z>0.6$, we require $\mathrm{FWHM}(\mathrm{Mg\,\textsc{ii}}) \geq 2500~\mathrm{km~s^{-1}}$. These criteria yield a sample of 333,908 broad-line AGNs.

We cross-match this sample with the FIRST catalog using a matching radius of $5^{\prime\prime}$, resulting in 12,639 radio-detected typical AGNs. We do not include Type~2 AGNs in the control sample, as only a small number of radio-detected Type~2 sources are available, and their radio emission is often dominated by extended star formation rather than nuclear jet activity. Excluding these sources ensures a physically meaningful comparison with the CL-AGN sample, which is dominated by nuclear radio emission.

\subsubsection{Radio Transient Sample}\label{Sec:radio_transient_sample}

Radio transients are commonly defined as sources that transition between radio-quiet and radio-loud states on timescales of years to decades. The classification is often based on the radio loudness parameter,
\begin{equation}
    R = \frac{f_{\rm 1.4\,GHz}}{f_{B}},
\end{equation}
where $f_{\rm 1.4\,GHz}$ and $f_{B}$ denote the rest-frame 1.4~GHz and $B$-band flux densities, respectively. Sources with $R<10$ are classified as radio-quiet, while those with $R>10$ are considered radio-loud \citep{1993ApJ...410...29S}.

In recent years, a growing population of radio transients has been reported, primarily identified as sources undetected in FIRST but detected in VLASS (or other newer radio surveys), consistent with rapid jet emergence or strong radio flaring associated with young/compact jets \citep{2020ApJ...905...74N,2021ApJ...914...22W,2022ApJ...938...43Z}. In this work, our comparison sample is compiled from the union of literature radio-transient catalogs, combining the sources reported by \citet{2020ApJ...905...74N}, \citet{2021ApJ...914...22W}, and \citet{2022ApJ...938...43Z}. This compilation contains in total $\gtrsim 50$ reported radio transients.
For a physically meaningful comparison to our CL-AGN sample, we further require that each transient has reliable measurements of redshift and the basic AGN properties needed in our analysis (e.g., black hole mass, accretion rate, and related quantities). After applying these quality cuts and removing objects lacking the required ancillary information, the final radio-transient comparison sample used throughout this paper consists of 22 sources.

\begin{figure}
    \centering
    \includegraphics[width=1\linewidth]{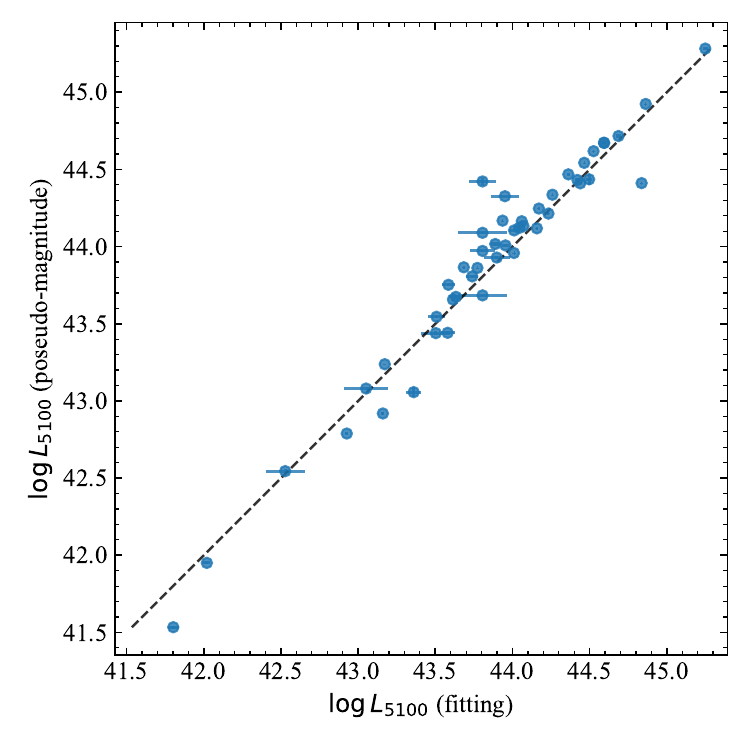}
    \caption{Comparison of the $L_{\rm 5100}$ luminosities derived from direct spectral fitting using {\tt PyQSOFit} and those inferred from the photometry-assisted filter–convolution method. The gray dashed line denotes the 1:1 line.}
    \label{Fig: L5100_L5100}
\end{figure}

\begin{figure}
    \centering
    \includegraphics[width=1\linewidth]{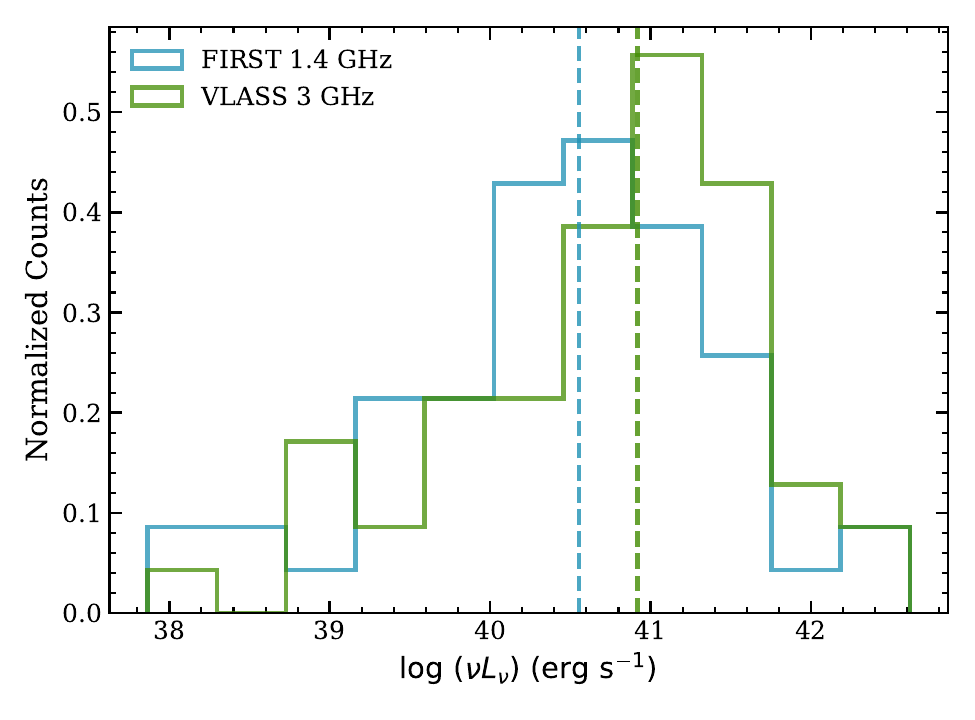}
    \caption{Normalized distribution of the CL-AGNs radio luminosity at $\rm 1.4GHz$ and $\rm 3GHz$. }
    \label{fig: L_radio}
\end{figure}

\section{Method}\label{Sec: method and result}


Irregular and rapid variability is a defining characteristic of AGNs and can strongly affect measurements of their radio-related properties, such as radio loudness and jet efficiency. Establishing optical properties that are contemporaneous with radio observations is therefore essential for investigating disk--jet coupling. However, optical spectroscopy is often sparsely sampled in time, and only a limited number of sources have spectroscopic observations obtained close to the epochs of radio measurements. 

To overcome this limitation, we combine optical spectroscopy with high-cadence photometric monitoring. In particular, the Zwicky Transient Facility (ZTF; \citealt{2019PASP..131a8002B,2019PASP..131a8003M}) provides multi-band optical photometry with a temporal baseline that overlaps well with the two released epochs of the VLASS radio survey. By anchoring photometric variability to spectroscopically derived continuum properties, we are able to infer optical luminosities that are approximately contemporaneous with the available radio observations.

\subsection{Spectral Analysis}\label{Sec: Spec_fitting}

Optical spectral decomposition provides the physical foundation for estimating black hole masses, isolating host-galaxy starlight, and establishing a physically motivated mapping between photometric fluxes and the intrinsic AGN continuum. We perform spectral fitting using the latest version of the \texttt{PyQSOFit} software,\footnote{\url{https://github.com/legolason/PyQSOFit}} which was originally developed in \textsc{IDL} by \citet{2019ApJS..241...34S} and later migrated to Python with further optimizations by \citet{2019MNRAS.482.3288G}. The most recent release incorporates a prior-informed principal component analysis (PCA) module that enables a robust separation of AGN and host-galaxy components \citep{2024ApJ...974..153R}.

Our optical spectroscopy is drawn primarily from SDSS and DESI, supplemented by a smaller number of spectra from LAMOST and other archival/follow-up observations when available. Many objects have multiple spectroscopic epochs spanning distinct accretion states. In our workflow, the faint-state spectrum (the epoch in which AGN signatures are weakest) is used to constrain the host-galaxy component, because the starlight fraction is highest and contamination from the variable AGN continuum is minimal. We then assume that the host contribution remains constant over the time span of our data and refit the bright-state spectra to measure the AGN continuum and broad emission-line properties needed for virial black hole mass estimates.

To ensure consistency among different spectroscopic epochs for each object, we place all spectra on a common relative flux scale. We adopt the narrow [$\rm O \textsc{iii}$] $\lambda5007$ emission line measured from the SDSS spectrum as the flux reference, as the narrow-line region extends over kiloparsec scales and its luminosity is expected to remain stable over decade-long timescales \citep{2002ApJ...574L.105B,2018MNRAS.477.4615D}. Spectra from DESI \citep{2023AJ....165....9S,2024AJ....168...95M}, LAMOST, and other sources are scaled accordingly.

During the fitting procedure, we first apply the PCA-based model to decompose the AGN and host-galaxy contributions. After subtracting the best-fit host component, we fit the line-free AGN continuum using a power-law combined with an $\rm Fe\,\textsc{ii}$ template. Emission-line complexes are then modeled using a single Gaussian for the narrow component and up to three Gaussians for the broad component to account for asymmetric profiles. Parameter uncertainties are estimated through 200 Monte Carlo realizations, in which random noise consistent with the flux uncertainties is added to the spectra.

We note that spectra from different facilities can have different instrumental resolutions, which in principle may affect detailed line-profile measurements. In this work, we do not explicitly homogenize the spectral resolution across surveys, because our primary use of the spectral fits is to robustly decompose and subtract the host-galaxy contribution and characterize broad-line widths at the order-of-magnitude level for virial black hole mass estimates. These quantities are dominated by the intrinsic widths of broad lines (typically $\gtrsim$ a few thousand $\rm km\,s^{-1}$) and are therefore only weakly affected by moderate differences in instrumental resolution. The impact of instrumental resolution is expected to be more important for narrow-line kinematics and subtle asymmetries, which we do not emphasize in the present work.

\subsection{Deriving $L_{\rm 5100}$ from Photometric Data}\label{Sec: L5100 from photometric}

To estimate optical continuum luminosities contemporaneous with the radio epochs, we infer $L_{\rm 5100}$ from near-simultaneous ZTF photometry. We select all ZTF measurements within $\pm50$ days of each VLASS observation and adopt their average magnitude as the contemporaneous optical brightness at time $t$. To ensure full coverage of the filter bandpass and to reduce contamination from strong emission-line variability, we adopt the ZTF $g$ band for sources at $z<0.3$ and the $r$ band for higher-redshift sources.

Assuming that the host-galaxy contribution remains constant over decade-long timescales, the observed-band flux can be decomposed as
\begin{equation}
F_{\rm total}=F_{\rm host}+F_{\rm emi}+F_{\rm cont},
\end{equation}
where $F_{\rm host}$, $F_{\rm emi}$, and $F_{\rm cont}$ denote the host-galaxy, emission-line, and AGN continuum contributions in the adopted photometric band, respectively. For each source, we use the spectroscopic epoch (denoted by subscript 0) to estimate $F_{\rm host}$ from the best-fit spectral decomposition (Section~\ref{Sec: Spec_fitting}). By convolving the best-fit total spectrum with the same filter response, we obtain the total flux $F_{\rm total,0}$ at the spectroscopic epoch, and thus the host-subtracted AGN-related flux $F_{\rm emi,0}+F_{\rm cont,0}=F_{\rm total,0}-F_{\rm host}$. At time $t$ near VLASS, the host-subtracted flux is similarly given by $F_{\rm emi,t}+F_{\rm cont,t}=F_{\rm total,t}-F_{\rm host}$.

To estimate the continuum contribution at time $t$, we adopt the empirical finding that continuum and emission-line luminosities vary with comparable amplitudes, $\Delta L_{\rm cont}:\Delta L_{\rm line}\approx 1:1$ \citep{2025ApJS..278...28G}, and assume that the change in host-subtracted flux between epochs 0 and $t$ is equally shared between the continuum and line components in the adopted band. This yields
\begin{equation}
F_{\rm cont,t} \approx F_{\rm cont,0} + \frac{\left[(F_{\rm total,t}-F_{\rm host})-(F_{\rm total,0}-F_{\rm host})\right]}{2},
\end{equation}
with $F_{\rm cont,0}$ taken from the spectral fit at the spectroscopic epoch.

We then convert $F_{\rm cont,t}$ to the rest-frame monochromatic luminosity at 5100~\AA\ by adopting a power-law continuum $f_{\lambda}\propto \lambda^{\alpha_{\rm O}}$ with $\alpha_{\rm O}=-1.5$. In practice, we normalize the power-law continuum such that its filter-convolved flux matches $F_{\rm cont,t}$, and evaluate the normalized continuum at rest-frame 5100~\AA\ to obtain $L_{\rm 5100}(t)$. Figure~\ref{Fig: L5100_L5100} compares the $\log L_{\rm 5100}$ values obtained directly from spectral fitting with those inferred from this photometric approach; the two estimates are consistent within the uncertainties, supporting the use of photometry as a proxy for $L_{\rm 5100}$ when contemporaneous spectroscopy is unavailable.

We note that throughout this work, $L_{\rm 5100}$ is primarily derived from ZTF photometry obtained close in time to the VLASS observations. An exception is made in Figure~\ref{fig:R_evolution}, where $L_{\rm 5100}$ used to compute the Eddington ratio is derived from contemporaneous spectral fitting to ensure a more reliable estimate of the accretion state when examining the evolution of radio loudness in CL-AGNs.

\section{Result and Discussion}\label{Sec: discussion}

\subsection{Origin of the Radio Emission}

Before investigating the physical coupling between the accretion flow and radio evolution, it is essential to establish the physical origin of the observed radio emission in our CL-AGN sample. 
Low-frequency radio emission in AGNs can arise from several distinct mechanisms, including relativistic jets, AGN-driven winds interacting with the circumnuclear medium, magnetically heated coronae, or host-galaxy star formation (SF) process (\citealt{2019NatAs...3..387P}).

\begin{figure*}
    \centering
    \includegraphics[width=\linewidth]{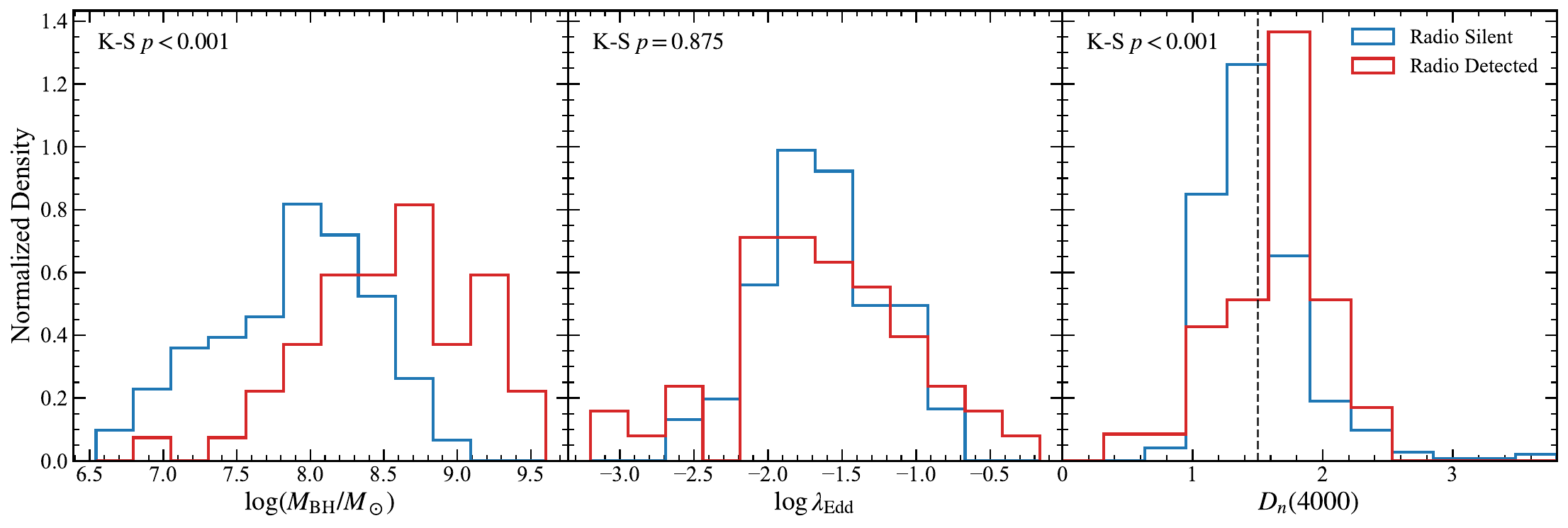}
    \caption{Normalized distributions of black hole mass (left), Eddington ratio (middle), and the $D_n(4000)$ index (right) for the radio-detected CL-AGNs (red) and the radio-silent CL-AGN control sample (blue). The $p$-values derived from two-sample K-S tests are indicated in each panel. The vertical dashed line in the right panel marks $D_n(4000) = 1.5$.}
    \label{fig: det_VS_sil}
\end{figure*}

To distinguish between nuclear AGN activity and host-galaxy star formation, we compare our 58 radio-detected sources with the control sample of 536 radio-silent CL-AGNs (described in Section \ref{Sec: radio silent sample}).
We examine their fundamental properties, $M_{\rm BH}$, $\lambda_{Edd}$, and $4000 \AA$ break index $D_{n}$(4000), as shown in Figure \ref{fig: det_VS_sil}.
The $D_{n}(4000)$ index is a highly reliable proxy for the age of a stellar population, with values $>1.5$ indicating quiescent galaxies with negligible recent star formation (\citealt{2003MNRAS.341...33K}).
A Kolmogorov-Smirnov (K-S) test reveals that the radio-detected CL-AGNs exhibit systematically larger $D_{n}(4000)$ values ($p<0.001$) than the radio-silent sample, with the vast majority peaking well above 1.5. Combined with their compact, unresolved radio morphologies in FIRST and VLASS, this supports the interpretation that their host galaxies are predominantly quiescent. The observed radio luminosities are therefore far too energetic to be sustained by star formation, strongly pointing to a nuclear AGN origin.

The comparison of accretion properties offers further physical insights into the CL mechanism. Interestingly, the two samples show statistically indistinguishable Eddington ratio distributions (K-S $p=0.875$), both strongly peaking at the transitional regime of $\lambda_{\rm Edd} \sim 0.01$.
This indicates that the critical threshold for CL transitions, widely interpreted as the state transition between a standard thin disk and a radiatively inefficient accretion flow (RIAF; e.g., \citealt{2023NatAs...7.1282R}), is a universal trigger for CL-AGNs, regardless of their radio properties.
In contrast, the radio-detected sample possesses systematically larger $M_{\rm BH}$ ($p<0.001$).
This difference is a natural consequence of the Fundamental Plane of black hole activity (\citealt{2003MNRAS.345.1057M}), which predicts that at a given accretion rate, more massive black holes produce intrinsically higher radio luminosities. Due to the flux limits of the FIRST and VLASS surveys, we are preferentially detecting the higher-mass end of the CL-AGN population whose radio output surpasses the detection threshold.

Regarding the specific nuclear mechanism, the radio loudness parameter ($\log R$) provides crucial diagnostic insights.
The vast majority of our radio-detected CL-AGNs have $\log R > 1$ (see Table \ref{tab: R_evolution} and Figure \ref{fig:R_evolution}), placing them unambiguously in the radio-loud (RL) regime. For these RL sources, the radio output is classically dominated by synchrotron radiation from relativistic jets.
However, a minor fraction of our sample falls into the radio-quiet (RQ) regime ($\log R < 1$). For these RQ outliers, the radio emission may not necessarily originate from a highly collimated jet; instead, it could be driven by non-thermal electrons accelerated within a magnetically heated corona (e.g., \citealt{2023ApJ...952L..28R, 2025A&A...699A..62J}) or by uncollimated disk winds.

Because the exact physical composition (jet, corona and wind) cannot be perfectly disentangled without high-resolution VLBI imaging, we hereafter adopt the broader term "radio activity" rather than strictly referring to "jet production". To quantify the non-thermal mechanical energy output of this nuclear radio activity, we utilize the empirical relation described in Equation \ref{Equ: Pj} (details shown in Section \ref{Sec: Jet power}). While this relation is primarily calibrated for estimating jet kinetic power in RL AGNs, and is therefore physically appropriate for the RL majority of our sample, we stress that for the RQ sources, the derived value ($P_{\rm j}$) should be interpreted strictly as an upper limit on the non-thermal mechanical power. By defining the ratio $P_{\rm j}/L_{\rm bol}$ as the radio-kinetic efficiency, we can robustly investigate the relationship between accretion states and radio activity across the entire population without over-interpreting the underlying emission mechanisms of individual borderline sources.

\subsection{Radio-kinetic Efficiency}\label{Sec: Jet power}

Out of a parent sample of 1092 CL-AGNs, we identify 58 sources with radio detections, corresponding to a detection rate of $5.31^{+1.49}_{-1.18}\%$. This occurrence rate is somewhat lower than the $\sim 13\%$ detection rate recently reported for CL-AGNs by \citet{2025arXiv250701355B}, and also lower than the historical FIRST detection fraction of $\sim 8\%$--$10\%$ for general optically selected broad-line quasars \citep[e.g.,][]{2002AJ....124.2364I, 2011ApJS..194...45S}. This apparent discrepancy is primarily driven by our highly conservative cross-matching criteria designed to maximize sample purity. Specifically, we enforce a strict $5''$ matching radius (reflecting the FIRST angular resolution) and a relatively high flux density threshold of $\geq 2$\,mJy, which filters out fainter or physically unassociated radio sources. Indeed, when applying these identical strict criteria to our control sample of typical AGNs drawn from SDSS DR14, the baseline detection rate is only $3.79^{+0.07}_{-0.06}\%$. Therefore, under a strictly controlled comparison, the CL-AGN population actually exhibits a slightly higher incidence of radio activity than typical AGNs. Among these 58 radio-detected CL-AGNs, 54 have measurements from both FIRST and VLASS, while the remaining four objects exhibit strong radio variability and are classified as radio transients, including two radio turn-on and two radio turn-off events (see Section~\ref{Sec: radio transients}).

We compute $k$-corrected monochromatic radio luminosities using
\begin{equation}
    \nu L_{\nu} = 4\pi D_{\rm L}^{2} \, \nu_{\rm rest} \, f_{\nu} (1+z)^{\alpha_{\rm R}-1},
\end{equation}
where $D_{\rm L}$ is the luminosity distance, $f_{\nu}$ is the observed-frame flux density at frequency $\nu_{\rm obs}$, $z$ is the redshift, and $\alpha_{\rm R}$ is the radio spectral index. The rest-frame frequency is given by $\nu_{\rm rest} = \nu_{\rm obs}(1+z)$.
We adopt a typical radio spectral index of $\alpha_{\rm R} = -0.7$ \citep{1969ApJ...155L..71K,1985Ap&SS.108..125A}. For FIRST and VLASS observations, we use reference frequencies of 1.4~GHz and 3.0~GHz, respectively.
With this definition, the resulting radio luminosities span $\log (\nu L_{\nu}) \sim 38$--$43~\mathrm{erg~s^{-1}}$, as shown in Figure~\ref{fig: L_radio}.

We further estimate the jet kinetic power following \citet{2017MNRAS.466.2294R},
\begin{equation}\label{Equ: Pj}
    P_{\rm j} = 5\times10^{22}\left(\frac{L_{\rm 1.4\,GHz}}{\mathrm{W\,Hz^{-1}}}\right)^{6/7}~\mathrm{erg~s^{-1}},
\end{equation}
where $L_{\rm 1.4\,GHz}$ is the rest-frame 1.4~GHz luminosity. The jet production efficiency is defined as $P_{\rm j}/L_{\rm bol}$, with the bolometric luminosity estimated as $L_{\rm bol}=9.26\,L_{\rm 5100}$ \citep{2006ApJS..166..470R}. This ratio provides an observational measure of the relative importance of mechanical jet power compared to radiative output.

\begin{figure}
    \centering
    \includegraphics[width=1\linewidth]{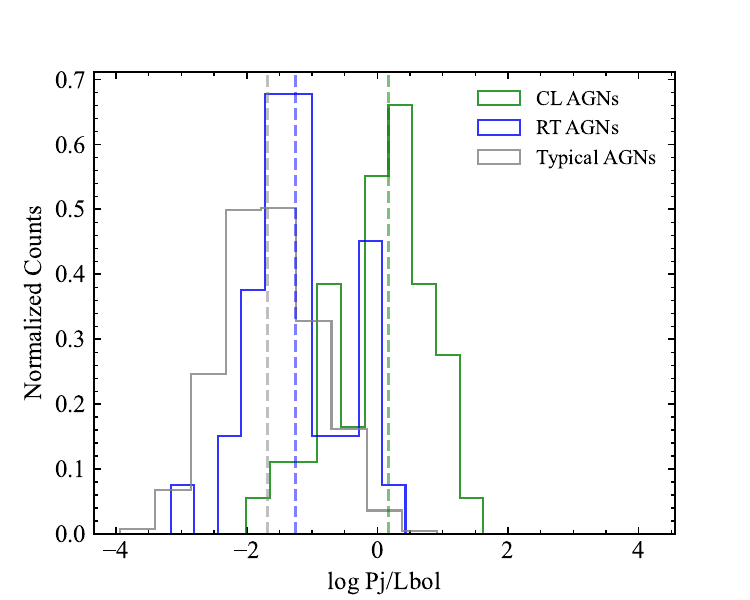}
    \caption{Normalized distribution of the jet production efficiency $P_{\rm j}/L_{\rm bol}$ for CL-AGNs (green), radio transients (blue), and typical AGNs (gray).}
    \label{fig: Pj/Lbol_distribution}
\end{figure}

Figure~\ref{fig: Pj/Lbol_distribution} presents the distribution of $P_{\rm j}/L_{\rm bol}$ for CL-AGNs, with typical AGNs and radio transients shown for comparison. CL-AGNs exhibit systematically higher radio-kinetic efficiencies than both control samples (K-S test: $p=3.44\times10^{-33}$ for CL-AGNs versus typical AGNs; $p=4.35\times10^{-8}$ for CL-AGNs versus radio transients). This enhancement is likely related to the tendency of CL-AGNs to be observed near the critical accretion regime associated with accretion-state transitions, a possibility that is explored further in Section~\ref{Sec: Pj/Lbol_Lbol/LEdd}.

\begin{figure*}
    \centering
    \includegraphics[width=1\linewidth]{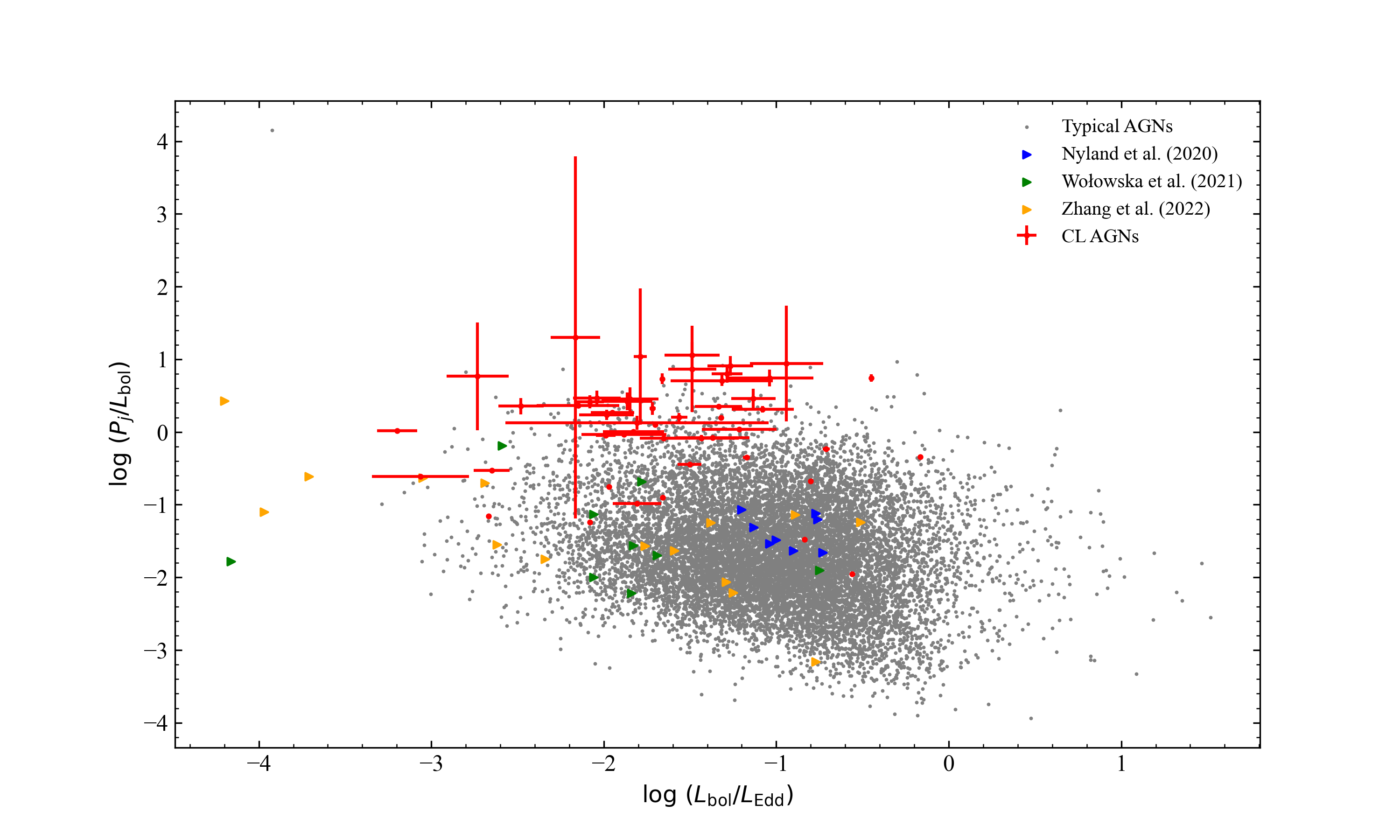}
    \caption{Distribution of CL-AGNs with radio detections (red dots) on the $P_{\rm j}/L_{\rm bol}$--$L_{\rm bol}/L_{\rm Edd}$ plane, shown together with a comparison sample of typical AGNs (gray circles) and radio transients (triangles). The typical AGN sample is drawn from the SDSS DR14 QSO catalog \citep{2020ApJS..249...17R}. The radio transient sample includes sources from \citet{2020ApJ...905...74N} (blue), \citet{2021ApJ...914...22W} (green) and \citet{2022ApJ...938...43Z} (orange), respectively.}
    \label{fig: Pj/L_REDD}
\end{figure*}

\subsection{Radio Radiation V.S. Eddington Ratio}\label{Sec: Pj/Lbol_Lbol/LEdd}

As noted earlier, radio emission in galaxies can arise from three primary components: star formation in host galaxy, AGN-driven outflows interacting with the interstellar medium, and the powerful relativistic jet activity (\citealt{2019NatAs...3..387P}). The 54 radio-detected CL-AGNs identified in our sample all exhibit compact radio morphologies, indicating that their radio output is dominated by the nuclear region rather than by star formation. Both AGN outflows and jet activity are closely linked to the accretion processes of the central SMBH, and radio radiation strength is generally expected to decline with increasing accretion rate (e.g., \citealt{2007ApJ...658..815S, 2017MNRAS.466.2294R}). Given that the prevailing interpretation of CL behavior invokes substantial changes in the accretion rate (e.g, \citealt{2017ApJ...846L...7S, 2018MNRAS.480.3898N, 2023ApJ...953...61Y, 2023NatAs...7.1282R}), CL-AGNs provide an ideal laboratory for investigating the connection between radio radiation and accretion state.
Recent time-domain radio studies of CL-AGNs further suggest a close connection between CL transition  and compact radio jets (e.g., \citealt{2023AcASn..64....7W, 2025arXiv250701355B}).

\subsubsection{Global Distribution}

As described in Section \ref{Sec: Spec_fitting}, we first fit the bright-state spectra of the CL-AGNs. We then estimate the black hole masses of the 54 CL-AGNs using the virial relation based on the luminosity and full width at half maximum (FWHM) of the broad $\rm H\beta$ emission line, following \citet{2005ApJ...627..721G}:
\begin{equation}
    \frac{M_{\rm BH}}{M_{\rm \sun}} = 3.6 \times 10^{6} \left( \frac{L_{\rm H\beta}}{10^{42} \rm erg\,s^{-1}} \right)^{0.56} \left( \frac{\rm FWHM_{H\beta}}{\rm 10^{3} km\,s^{-1}} \right)^{2.06}.
\end{equation}
Combining these black hole mass estimates with the contemporaneous $L_{\rm 5100}$ 
values derived from photometric data (Section \ref{Sec: L5100 from photometric}), we further compute the Eddington ratio for each source followed \cite{2006ApJS..166..470R}:
\begin{equation}
\lambda_{\rm Edd} \equiv \frac{L_{\rm bol}}{L_{\rm Edd}}
= \frac{9.26L_{5100}}{1.38 \times 10^{38}(M_{\rm BH}/M_{\odot})} .
\end{equation}

Figure \ref{fig: Pj/L_REDD} shows the distribution of the radio-kinetic efficiency ($P_{\rm j}/L_{\rm bol}$) as a function of the Eddington ratio ($L_{\rm bol}/L_{\rm Edd}$). Both radio transients and typical AGNs exhibit a clear trend of decreasing radio-kinetic efficiency with increasing accretion rate. In contrast, this relation appears significantly more scattered for the CL-AGN population. This increased dispersion is likely driven by the relatively small sample size and the narrow range of Eddington ratios spanned by CL-AGNs, limiting the dynamic range over which the trend can be examined. 
Nevertheless, CL-AGNs systematically show higher radio output than both radio transients and typical AGNs, effectively populating the low-accretion-rate regime of typical AGNs and broadly extending the global $P_{\rm j}/L_{\rm bol}$--$L_{\rm bol}/L_{\rm Edd}$ relation. 
To quantitatively confirm this, we calculated the accretion rate properties of the two populations. The 54 radio-detected CL-AGNs in our sample exhibit a median logarithmic Eddington ratio of $-1.66$, which is systematically lower than the median value of $-1.01$ for the typical SDSS AGN sample. A two-sample K-S test confirms that the accretion rates of CL-AGNs are statistically distinct and significantly lower (statistic = $0.43$, $p = 6.02 \times 10^{-9}$). 
We note that these Eddington ratios are derived from the bright-state spectra associated with the CL transitions rather than from arbitrary epochs. Therefore, they characterize the accretion states immediately before or after the CL events, depending on whether the source is classified as turn-off or turn-on.

This characteristic clustering around $\lambda_{\rm Edd} \sim 0.01$ is profoundly significant. From a theoretical perspective, this value corresponds to the critical threshold where the accretion flow undergoes a fundamental state transition between a radiatively efficient standard thin disk and a RIAF \citep[ e.g.,][]{2023NatAs...7.1282R}.
Because the CL phenomenon is thought to be preferentially associated with accretion flows operating near the critical accretion threshold \citep[e.g.,][]{2024ApJ...966..128W, 2025RAA....25i5012C}, their enhanced radio-kinetic efficiency is physically well-justified. It is broadly consistent with the negative correlation between non-thermal radio output and accretion rate expected in accreting black hole systems \citep[e.g.,][]{2007ApJ...658..815S, 2021ApJ...914...22W, 2022ApJ...938...43Z}. However, given the large intrinsic scatter observed within the CL-AGN sample, whether they individually follow the exact same declining trend of radio output with increasing accretion rate remains to be tested with larger, high-cadence tracking samples.


\subsubsection{Source-by-source Behavior}

\begin{figure}
    \centering
    \includegraphics[width=1\linewidth]{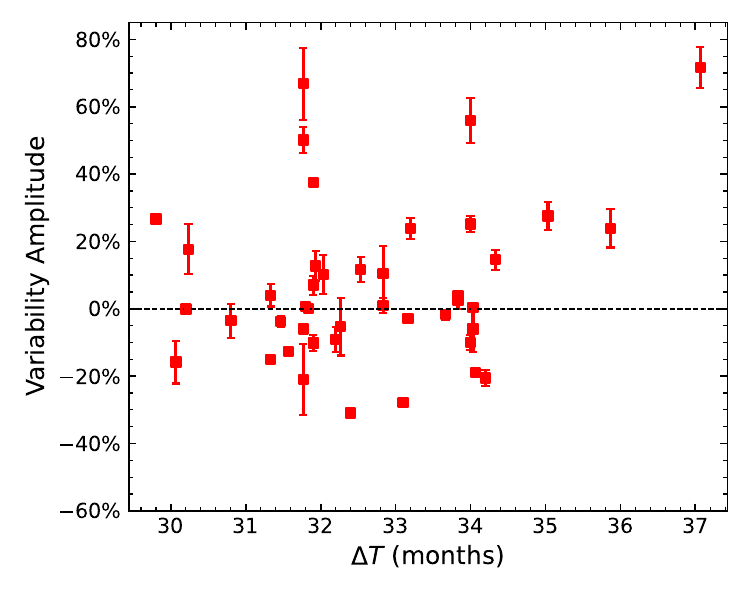}
    \caption{Fractional radio flux density variations of CL-AGNs between the two VLASS epochs, defined as $(F_{\rm epoch 2}-F_{\rm epoch 1})/(F_{\rm epoch 1})$.}
    \label{fig: radio_vari}
\end{figure}

For the majority of AGNs, radio flux densities vary only mildly on long timescales (\citealt{1992ARA&A..30..575C, 2005ApJ...618..108B}). The CL-AGN population appears to follow a similar behavior. Figure \ref{fig: radio_vari} shows the fractional change in radio flux density, $(F_{\rm epoch 2}-F_{\rm epoch 1})/F_{\rm epoch 1}$, between the two VLASS epochs (epoch 1 and epoch 2) for our CL-AGN sample. We find that the variations are typically within $\sim$20\%, indicating that the radio flux densities of CL-AGNs can be regarded as approximately constant over a timescale of $\sim$3 years.

\begin{figure}
    \centering
    \includegraphics[width=1\linewidth]{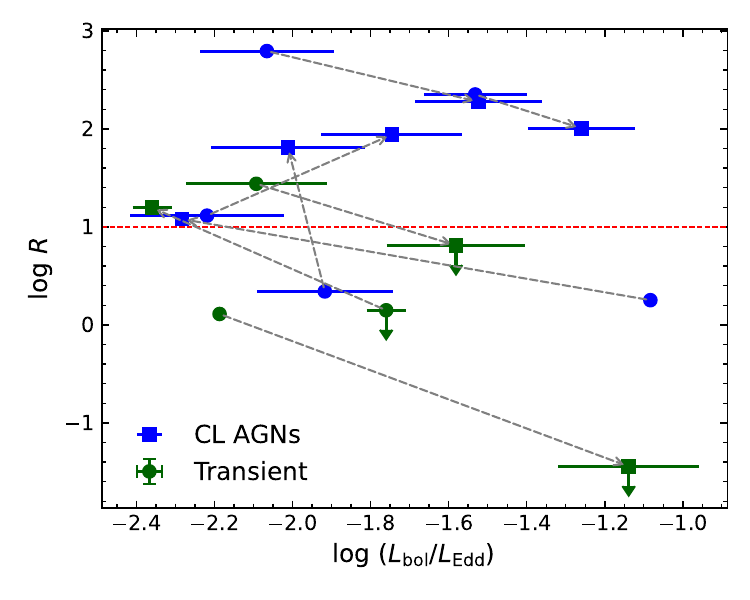}
    \caption{The distribution of CL-AGNs detected in both the FIRST and VLASS surveys (blue symbols) and radio transients identified among CL-AGNs (green symbols) on the $\log R$–$\log (L_{\rm bol}/L_{\rm Edd})$ plane. The gray dashed arrows indicate the evolutionary trajectories of CL-AGNs from the FIRST epoch to the VLASS epoch. The red dashed line marks $\log R = 1$, commonly adopted as the boundary between radio-loud and radio-quiet sources.}
    \label{fig:R_evolution}
\end{figure}

Motivated by this result, we adopt a matching window of 1000 days to select sources with quasi-simultaneous optical spectroscopic observations and both FIRST and VLASS radio measurements, enabling us to investigate the evolution of radio properties in CL-AGNs over a timescale of $\sim$20 years.
We identify five sources with such contemporaneous multi-epoch observations are shown in Table \ref{tab: R_evolution}. As shown in Figure \ref{fig:R_evolution}, on the $R$–$L_{\rm bol}/L_{\rm Edd}$ plane, all of these sources exhibit a trend of decreasing radio emission with increasing accretion rate, with the exception of J113615.08-002314.2 (the detail is shown in Section \ref{Sec: TDE_flare}). 
Although a strong anticorrelation is present for each individual source, the slopes are not consistent across the sample. This diversity likely reflects source-specific behaviors, such as the cumulative or time-integrated nature of radio emission. Nevertheless, the behavior observed in CL-AGNs provides compelling evidence for an overall negative correlation between radio emission and accretion rate, supporting the presence of disk–jet coupling.

\begin{deluxetable*}{cccccccccc}[!ht]
    \setlength{\tabcolsep}{2pt}
    \renewcommand{\arraystretch}{1.2}
    \tablenum{1}
    \centering
    \tablecaption{The information of the contemporaneous muti-epoch observational CL-AGNs and radio transients. \label{tab: R_evolution}}
    \tablehead{\colhead{Object Name} & \colhead{$f_{\rm peak, FIRST}$} & \colhead{$f_{\rm int, FIRST}$} & \colhead{$\rm RMS_{\rm FIRST}$} & \colhead{$f_{\rm peak, VLASS}$} & \colhead{$f_{\rm int, VLASS}$} & \colhead{$\log R_{\rm FIRST}$} & \colhead{$\log R_{\rm VLASS}$} & \colhead{$\lambda_{\rm EDD, FIRST}$} & \colhead{$\lambda_{\rm EDD, VLASS}$} \\
    \colhead{(1)} & \colhead{(mJy, 2)} & \colhead{(mJy, 3)} & \colhead{(mJy, 4)} & \colhead{(mJy, 5)} & \colhead{(mJy, 6)} & \colhead{(7)} & \colhead{(8)} & \colhead{(9)} & \colhead{(10)} }
    \startdata
J023823.27-061619.9 & 63.08 & 80.75 & 0.13 & $19.41 \pm 0.21$ & $97.25 \pm 1.67$ & 2.79 & 2.28 & $-2.07 \pm 0.17$ & $-1.52 \pm 0.16$ \\
J132045.25-002449.6 & 3.88 & 4.84 & 0.14 & $6.08 \pm 0.20$ & $6.67 \pm 0.37$ & 0.34 & 1.80 & $-1.92 \pm 0.18$ & $-2.01 \pm 0.20$ \\
J113615.08-002314.3 & 5.97 & 6.34 & 0.17 & $11.69 \pm 0.19$ & $10.63 \pm 0.31$ & 1.11 & 1.82 & $-2.22 \pm 0.20$ & $-1.75 \pm 0.18$ \\
J021359.79+004226.6 & 3.46 & 3.05 & 0.11 & $1.72 \pm 0.15$ & $1.91 \pm 0.29$ & 0.25 & 1.08 & -1.08 & -2.28 \\
J001553.74+040039.5 & 9.07 & 7.98 & 0.13 & $11.12 \pm 0.21$ & $11.40 \pm 0.37$ & 2.35 & 2.00 & $-1.53 \pm 0.13$ & $-1.26 \pm 0.14$ \\
J230443.60-084110.0 & 17.69 & 22.48 & 0.16 & \textless{}0.70 & \textless{}0.70 & 0.11 & $<-1.45$ & $-2.19 \pm 0.00$ & $-1.14 \pm 0.18$ \\
J092313.53+043445.0 & 2.57 & 3.95 & 0.15 & \textless{}0.70 & \textless{}0.70 & 1.44 & $<0.81$ & $-2.09 \pm 0.18$ & $-1.58 \pm 0.18$ \\
J094144.83+575123.8 & \textless{}1.00 & \textless{}1.00 & - & $3.40 \pm 0.17$ & $4.71 \pm 0.36$ & $<0.15$ & 1.19 & $-1.76 \pm 0.05$ & $-2.36 \pm 0.05$ \\
J102752.40+421012.4 & \textless{}1.00 & \textless{}1.00 & - & $4.61 \pm 0.12$ & $4.81 \pm 0.21$ & - & - & - & -
\enddata
    \tablecomments{Column (1): object name;
Columns (2--6): radio flux densities and their uncertainties measured from FIRST and VLASS;
Columns (7--8): radio loudness parameters, calculated as $\log R = \log (f_{\rm 1.4\,GHz}/f_{\rm B}$) in the rest frame;
Columns (9--10): Eddington ratios estimated at the epochs of the FIRST and VLASS observations.
}
\end{deluxetable*}

\subsubsection{J113615.08--002314.2: A Pre-VLASS Flare and Possible Outflow Activity}\label{Sec: TDE_flare}

J113615.08--002314.2 exhibits a radio evolution distinct from that of the other sources in Figure~\ref{fig:R_evolution}, showing an enhancement of radio emission coincident with an increase in the inferred accretion rate. This behavior contrasts with the general trend observed in the rest of the sample and suggests the presence of an additional physical process affecting the radio output.

\begin{figure}
    \centering
    \includegraphics[width=1 \linewidth]{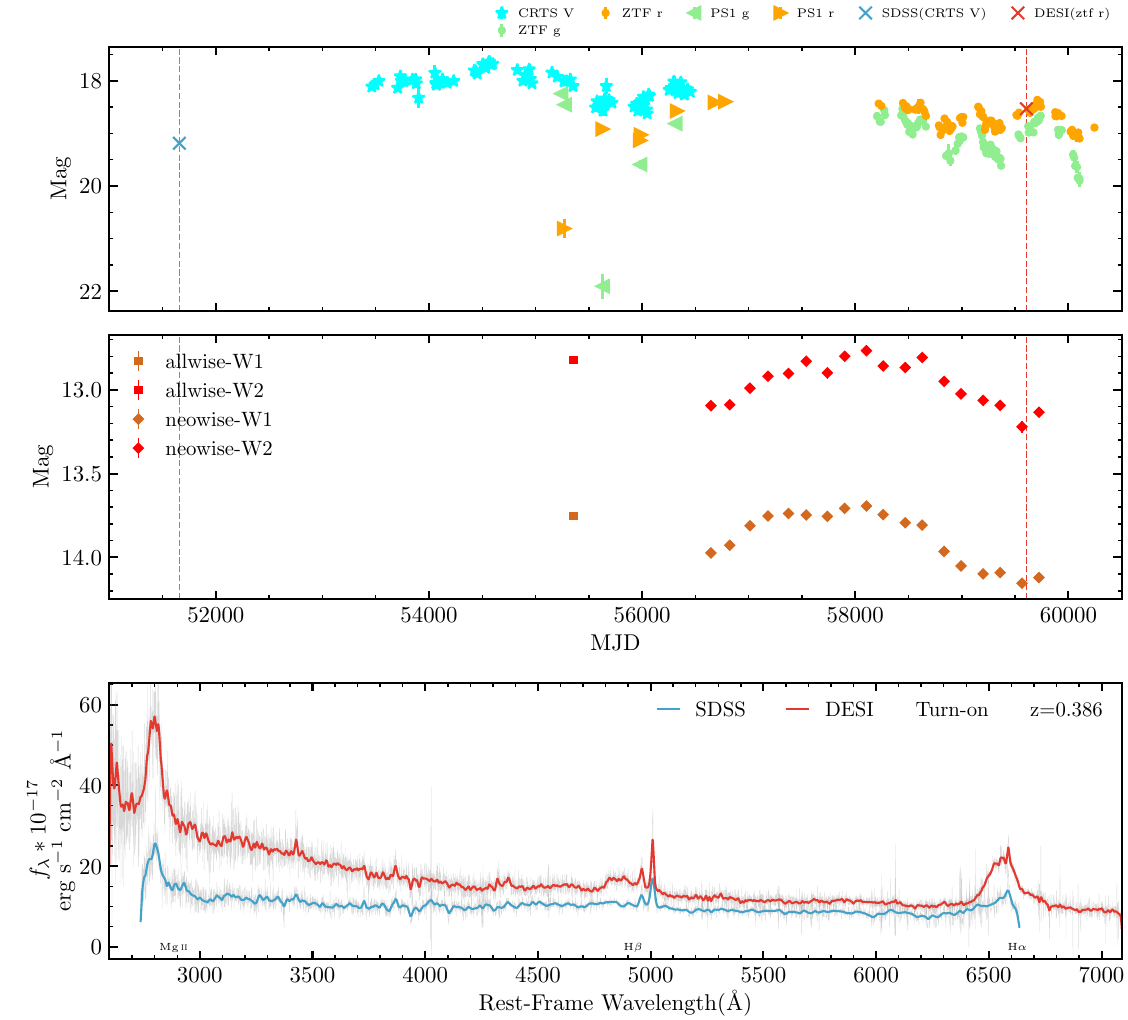}
    \caption{The light curves and spectra for J113615.08-002314.2. Top panel and middle panel show the optical and mid-infrared light curves. The bottom panel show the  SDSS and DESI spectra.}
    \label{fig: J1136}
\end{figure}

The multiwavelength light curves indicate that this source experienced a significant flare prior to the VLASS observations (Figure~\ref{fig: J1136}). The Pan-STARRS1 $g$-band data show a substantial brightening around MJD $\sim$ 55000, with an amplitude of nearly four magnitudes. However, the sparse temporal coverage, together with the elevated flux levels seen in both the earlier CRTS and later ZTF observations, introduces considerable uncertainty regarding the detailed evolution and significance of the optical flare.
In contrast, the mid-infrared variability observed in the \textit{NEOWISE} $W1$ and $W2$ band is more clearly defined, exhibiting a rapid rise beginning before MJD $\sim$ 56050, a peak around MJD $\sim$ 58000, and a subsequent gradual decline. 
Such mid-infrared flare profiles bear qualitative resemblance to those reported for TDE candidates \citep[e.g.,][]{2021ApJS..252...32J}, potentially reflecting dust reprocessing of an earlier energetic episode in the nuclear region.
Despite this resemblance, no significant variability is detected in the $\rm He\,\textsc{ii}$ emission line in the DESI spectrum, a feature commonly associated with many confirmed TDEs \citep[e.g.,][]{2019ApJ...887..218L}. This may indicate that the spectroscopic observation was obtained after the flare had subsided, or that any line variability was intrinsically weak and below the detection threshold. Overall, the available data do not provide sufficient evidence to securely classify this event as a TDE.

Nevertheless, energetic nuclear flares, regardless of their origin, are expected to drive substantial mass outflows, as a significant fraction of the liberated material may be expelled rather than accreted onto the central black hole \citep{2020MNRAS.499..482N, 2023AN....34430015K}. 
Interaction between such outflows and the surrounding circumnuclear or interstellar medium can naturally lead to delayed, enhanced radio emission. Such late-time radio brightening is a well-documented feature in TDEs, where radio emission often emerges or peaks months to years after the initial optical/X-ray flare as the newly launched outflow decelerates \citep{2020SSRv..216...81A}.

Moreover, recent high-resolution studies of extreme CL-AGNs have revealed a profound, time-delayed connection between accretion/coronal disruptions and radio activity. For instance, in the iconic CL-AGN 1ES 1927+654, the sudden destruction of the X-ray corona was accompanied by an immediate radio dip \citep{2022ApJ...931....5L}, strongly linking the inner coronal magnetic fields to the radio output. Remarkably, a spatially resolved radio jet was discovered in this same source approximately three years after the initial CL event \citep{2025ApJ...981..125L, 2025ApJ...979L...2M}, demonstrating that catastrophic accretion state transitions can indeed launch new jets or outflows that become radio-bright at late times.

In this context, the pre-VLASS flare in J113615.08--002314.2 likely triggered a similar sequence: a violent accretion episode that subsequently launched an outflow or compact jet. Given the required time delay for the outflow to expand and shock the ambient medium, the enhanced radio luminosity observed in the later VLASS epoch naturally explains the apparent positive correlation between the radio emission and the elevated accretion rate. While the flare properties share similarities with TDEs, we emphasize that the current data do not uniquely support a pure TDE interpretation. Instead, this source likely represents a broader class of transient nuclear activity, either a TDE or a violent accretion instability akin to 1ES 1927+654, where a prior energetic episode governs the delayed radio behavior, highlighting the diverse pathways that can produce radio enhancement.

\subsection{Radio Transients in CL-AGNs}\label{Sec: radio transients}


Radio variability is generally modest in the majority of AGNs, especially on timescales of years to decades. However, a small subset of AGNs exhibits dramatic changes in radio emission, commonly referred to as radio transients. The identification of such sources provides a unique opportunity to probe rapid changes in radio activity and their connection to accretion processes. 

Previously reported radio transients are predominantly radio turn-on sources, namely objects that are undetected in FIRST but become detectable in VLASS. These sources are generally interpreted as systems that have recently launched compact, small-scale jets (\citealt{2020ApJ...905...74N, 2021ApJ...914...22W, 2022ApJ...938...43Z}). In principle, radio turn-off sources are also expected, in which the jet may be approaching the end of its lifetime or the outflow activity has ceased. However, such sources have rarely been reported. This is partly due to their intrinsic rarity, and partly because the radio spectral index is typically negative, such that sources detected in FIRST but absent from the VLASS catalog may simply result from their steep radio spectra rather than genuine fading of radio emission.

Furthermore, interpreting radio variability and spectral shapes in the $\sim$ 100s of MHz to 45 GHz regime requires extreme caution, as this frequency range represents a physically complex zone governed by an interplay of various emission and absorption processes \citep[e.g.,][]{2019NatAs...3..387P}. For compact, newly launched jets commonly associated with radio turn-on transients, the radio spectra are often peaked or inverted at low frequencies due to strong local absorption. Two primary absorption mechanisms dictate the spectral shape in this regime: synchrotron self-absorption (SSA) sets in at $<45$\,GHz if the magnetic field within the compact region is sufficiently high, whereas free-free absorption (FFA) becomes dominant if the viewing angle passes through dense, ionized thick clouds \citep{2025FrASS..1130392L}. In the context of CL-AGNs, such ionized thick clouds could naturally arise from massive outflows expelled during the violent accretion state transitions. 
According to \cite{1966AuJPh..19..195K}, the spectral index expected from pure synchrotron self-absorption is theoretically limited to $\alpha_{\rm R} \leq 2.5$. Therefore, sources exhibiting extremely inverted spectra with $\alpha_{\rm R} > 2.5$ are more likely to reflect genuine intrinsic variations in radio flux or the presence of strong FFA, rather than being dominated solely by SSA effects. 
However, radio emission enhanced by outflow-driven shocks is generally expected to be accompanied by an increase in the accretion rate (e.g., the case of J113615.08–002314.2 discussed in Section \ref{Sec: TDE_flare}; \citealt{2022ApJ...938...87K}). In contrast, the radio transients identified in our CL-AGN sample consistently exhibit a declining accretion rate alongside enhanced radio emission. This discrepancy suggests that the observed radio variability is unlikely to be dominated by shock-powered emission, and is instead more plausibly attributed to compact, unresolved jet activity.

In this work, we identify four radio transients within the CL-AGN sample, including two radio turn-on and two radio turn-off sources, as summarized in Table \ref{tab: R_evolution}.
We place these radio transients on the $R$–$L_{\rm bol}/L_{\rm Edd}$ plane (excluding J102752.40+421012.4, for which spectroscopic data are unavailable), as shown in Figure \ref{fig:R_evolution}. These peculiar CL-AGNs continue to exhibit an anticorrelation between radio radiation strength and accretion rate.
On the other hand, by analogy with X-ray binaries, an accretion rate of $L_{\rm bol}/L_{\rm Edd} \sim 0.01$ is often regarded as a critical threshold for state transitions in CL-AGNs, at which the accretion flow is expected to evolve from a standard thin disk to a radiatively inefficient accretion flow (e.g., \citealt{2019ApJ...883...76R, 2020MNRAS.492.2335L}). As illustrated in Figure \ref{fig:R_evolution}, the accretion-rate variations of these radio transients appear larger than those of the remaining CL-AGNs, and all of them span this critical accretion regime. This behavior provides further evidence for a coupling between the accretion disk and jet activity.

\begin{figure}
    \centering
    \includegraphics[width=1\linewidth]{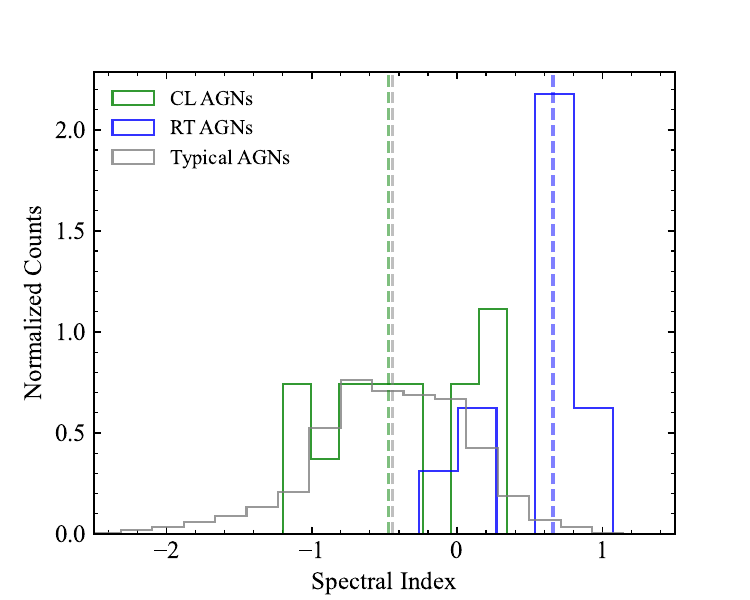}
    \caption{Normalized distribution of radio spectral index for CL-AGNs (green), radio transients (blue), and typical AGNs (gray).}
    \label{fig: radio spectral index}
\end{figure}

In most AGNs, the radio emission is dominated by optically thin synchrotron radiation from extended, long-lived jets, resulting in a steep radio spectrum with a negative spectral index. In contrast, radio transients are often associated with newly launched or rapidly evolving compact jets, where synchrotron self-absorption becomes important at low frequencies. This leads to a peaked radio spectrum, characterized by a positive spectral index below the turnover frequency. Therefore, several previous studies have employed inverted or rising radio spectra as an efficient criterion to identify radio transients and young jet activity in AGNs (e.g., \citealt{2025ApJ...987..170C}).
Here, we cross-match the radio-detected CL-AGNs with the LOTSS DR2 catalog at a central frequency of 0.144 GHz. Because LOFAR and VLASS observations are quasi-simultaneous, this approach minimizes the impact of radio flux variability. In total, we identify 16 CL-AGNs with detections in both LOFAR and VLASS  surveys. For these sources, we derive the radio spectral index $\alpha_{\rm R}$, and compare their distributions with those of typical AGNs and previously reported radio transients, as shown in Figure \ref{fig: radio spectral index}.
It is evident that the radio transients almost exclusively exhibit positive spectral indices, whereas the vast majority of CL-AGNs and typical AGNs show negative $\alpha_{\rm R}$ values. A Kolmogorov–Smirnov test further indicates no statistically significant difference between the $\alpha_{\rm R}$ distributions of CL-AGNs and typical AGNs (Kolmogorov-Smirnov test: $p=5.384 \times 10^{-04}$ for CL-AGNs versus radio transients; $p=7.948 \times 10^{-1}$ for CL-AGNs versus typical AGNs). This suggests that, for most CL-AGNs, the radio emission is produced by relatively stable jets, similar to those in typical AGNs, while the four radio transients identified in the CL-AGN sample are likely rare or stochastic events.

\subsection{Connection between CL Behavior and Radio Variability}\label{Sec:radio delay}

Previous studies have established that a coupling between the accretion disk and relativistic jets is a common feature of accreting black hole systems, spanning from X-ray binaries to active galactic nuclei (e.g., \citealt{2003MNRAS.345.1057M, 2004MNRAS.355.1105F}). In this framework, changes in accretion state are often accompanied by systematic variations in jet activity (\citealt{2006MNRAS.372.1366K, 2008ARA&A..46..475H}). Whether the CL phenomenon in AGNs can directly trigger a corresponding response in jet activity, however, remains observationally challenging to establish.
An important caveat in addressing this question is that the radio emission probed by surveys such as FIRST and VLASS predominantly traces kpc-scale structures, including extended jets and lobes. In contrast, CL transitions occur on relatively short timescales (years to decades) and are physically associated with changes in the inner accretion flow, which can only influence the pc-scale jet or compact radio core. As a result, any jet response directly linked to individual CL events is expected to manifest primarily on pc scales, while the kpc-scale radio emission may remain largely insensitive to such rapid accretion-state changes.

To further establish a causal connection, it is crucial to consider the temporal sequence of the radio observations relative to the CL events. 
Figure~\ref{fig:timeline} presents the observation timelines for all 58 radio-detected CL-AGNs, showing the epochs of optical spectroscopic observations together with the available radio measurements from FIRST and VLASS.
The FIRST observations (1993--2011) generally represent the earliest radio baseline, while the VLASS observations (2017--2022) probe the later radio state, often contemporaneous with or preceding recent optical spectroscopy from surveys such as DESI. Exact observation dates and radio flux measurements are provided in Appendix Table~\ref{tab:radio_clagn}.
As illustrated by the timelines, the relative temporal ordering between the radio observations and the CL transitions varies substantially from source to source. In many cases, the radio observations bracket the spectroscopically identified CL event, while in others the available radio epochs occur entirely before or after the transition. Consequently, assigning a strict ``before'' and ``after'' interpretation to the radio measurements is limited by the sparse temporal sampling available for most sources. Over the $\sim20$ yr interval between FIRST and VLASS, an AGN may undergo multiple unrecorded CL episodes that are not captured by existing optical spectroscopy.
Furthermore, even when a radio observation follows a documented CL transition, a radio response is not necessarily expected immediately because jet formation and propagation may introduce significant physical delays. Recent high-cadence monitoring of the extreme CL-AGN 1ES~1927+654 demonstrates this explicitly: while the destruction of the X-ray corona produced an immediate radio dimming \citep{2022ApJ...931....5L}, the emergence of a newly launched, spatially resolved radio jet was not detected until approximately three years after the initial event \citep{2025ApJ...981..125L,2025ApJ...979L...2M}. Therefore, interpreting radio variability in CL-AGNs requires accounting for both the possibility of multiple intervening accretion-state transitions and the finite timescales associated with the launching and propagation of radio-emitting outflows.


On the one hand, our results based on the continuous evolution of CL-AGNs show a clear tendency for radio emission to weaken with increasing Eddington ratio. It is worth noting, however, that the radio strength adopted in this work is defined relative to the optical luminosity. During the bright states of CL-AGNs, an increase in optical luminosity alone can naturally lead to a decrease in the relative radio loudness, even if the absolute radio luminosity remains unchanged. Indeed, for a large fraction of CL-AGNs, we do not observe significant variations in their absolute radio luminosities (e.g., Figure 5). This behavior is consistent with expectations if the observed radio emission is dominated by kpc-scale components, whose radiative evolution occurs on much longer timescales than the optical variability associated with CL transitions. Any prompt radio response confined to the pc-scale core could therefore be diluted, delayed, or smeared out when measured with arcsecond-resolution radio observations spanning two decades.

On the other hand, we identify a small number of radio transients within the CL-AGN population that exhibit behavior broadly consistent with the disk--jet coupling scenario. In these cases, compact or small-scale jet activity may be temporarily enhanced as AGNs transition toward low-accretion states, followed by a decline in radio emission as the accretion rate increases again. However, the radio spectral index distribution indicates that the majority of CL-AGNs do not share the same radio properties as these transients, suggesting that they do not constitute a homogeneous population. Instead, such radio transients are likely rare or stochastic events, rather than a universal consequence of the CL phenomenon.

Taken together, while our results support a general anti-correlation between radio emission and accretion rate on kpc scales, they do not yet provide definitive evidence that CL transitions directly regulate jet activity on the relevant physical scales. A more direct test of the CL--jet connection will require high-angular-resolution radio observations, such as VLBI, capable of isolating pc-scale radio emission, as well as coordinated, multi-wavelength monitoring to simultaneously trace accretion-driven variability and jet evolution.

\section{Summary}\label{Sec: Summary}

In this work, we investigate the long-term radio properties of changing-look active galactic nuclei (CL-AGNs) over a timescale of approximately 20 years by combining the wide-area FIRST and VLASS radio surveys. By incorporating optical spectroscopic constraints and high-cadence photometric monitoring that are quasi-simultaneous with the radio observations, CL-AGNs provide a unique laboratory for testing the coupling between accretion flows and relativistic jets in supermassive black holes. Our main results can be summarized as follows.

\begin{enumerate}
\item We find that the radio detection fraction of CL-AGNs is $\sim$5.31\%, slightly higher than that of typical AGNs. Moreover, radio-detected CL-AGNs exhibit systematically higher jet production efficiencies, quantified by $P_{\rm j}/L_{\rm bol}$, than both typical AGNs and radio transients. This enhancement is consistent with the tendency of CL-AGNs to preferentially occupy the low-accretion-rate regime.

\item At the population level, the expected anti-correlation between radio emission strength and accretion rate is weak for CL-AGNs as a whole, in contrast to the clearer trends observed in typical AGNs and radio transients. However, when focusing on a small subset of CL-AGNs with continuous multi-epoch coverage over the $\sim$20-year baseline, a clear source-by-source anti-correlation emerges, indicating that disk--jet coupling becomes evident when long-term evolutionary behavior is resolved.

\item We identify a notable counterexample in which both the accretion rate and radio emission increase simultaneously. Based on its infrared variability and emission-line properties, this source shows signatures that may be associated with tidal disruption event–like activity, raising the possibility that transient accretion episodes could contribute to enhanced radio emission.

\item We report the first identification of four radio transients within a CL-AGN sample. These objects exhibit rapid radio turn-on or turn-off behavior on decadal timescales, consistent with the emergence or disappearance of compact jets. However, such radio transient CL-AGNs represent rare and exceptional cases and do not characterize the CL-AGN population as a whole.

\item Taken together, our results indicate that disk--jet coupling in CL-AGNs is governed primarily by long-term accretion history and jet evolution rather than by instantaneous accretion state changes. Whether changing-look transitions directly trigger or suppress jet activity remains uncertain and will require future coordinated, multi-wavelength monitoring to establish a definitive causal connection.
\end{enumerate}

\section*{acknowledgements}

We acknowledge the anonymous referee for valuable comments that helped to improve the paper.
This research is supported by the National Key R\&D Program of China with grant No. 2023YFA1608100. This work is supported by the National Natural Science Foundation of China (NSFC) under grant No. 12503019 and 12273013. The authors also acknowledge support from the National Key R\&D Program of China (grant Nos. 2023YFA1607804, 2022YFA1602902, 2023YFA1607800), other NSFC projects (grant Nos. 12120101003, 12373010, 12173051, 12233008, 12403022, and 12103048), and the China Manned Space Project (No. CMS-CSST-2025-A06). The authors also acknowledge the Strategic Priority Research Program of the Chinese Academy of Sciences with grant Nos. XDB0550100 and XDB0550000.
MFG is supported by the National Science Foundation of China (grant 12473019), the Shanghai Pilot Program for Basic Research-Chinese Academy of Science, Shanghai Branch (JCYJ-SHFY-2021-013), the National SKA Program of China (Grant No. 2022SKA0120102), and the China Manned Space Project with No. CMS-CSST-2025-A07.
Y.M.C. acknowledges support by the National Natural Science Foundation of China, NSFC grant No. 12333002.

We acknowledge the use of data from the Faint Images of the Radio Sky at Twenty Centimeters survey and the Very Large Array Sky Survey. 
The National Radio Astronomy Observatory is a facility of the National Science Foundation operated under cooperative agreement by Associated Universities, Inc.
This work also makes use of data from the Sloan Digital Sky Survey (SDSS) and the Dark Energy Spectroscopic Instrument (DESI). 
Funding for SDSS has been provided by the Alfred P. Sloan Foundation, the U.S. Department of Energy Office of Science, and the Participating Institutions. 
DESI is managed by the U.S. Department of Energy's Lawrence Berkeley National Laboratory with primary funding from the U.S. Department of Energy Office of Science and the National Science Foundation.
Optical time-domain data were obtained from the Zwicky Transient Facility, which is supported by the National Science Foundation under grant No. AST-2034437.

\appendix

\section{Radio-detected CL-AGN Catalog and Observation Timeline}

Table~\ref{tab:radio_clagn} presents the complete catalog of the 58 radio-detected CL-AGNs analyzed in this work, including their redshifts, optical spectroscopic epochs, radio flux densities from FIRST and VLASS, and derived physical parameters. 

\begin{figure*}
    \centering
    \includegraphics[width=0.98\linewidth]{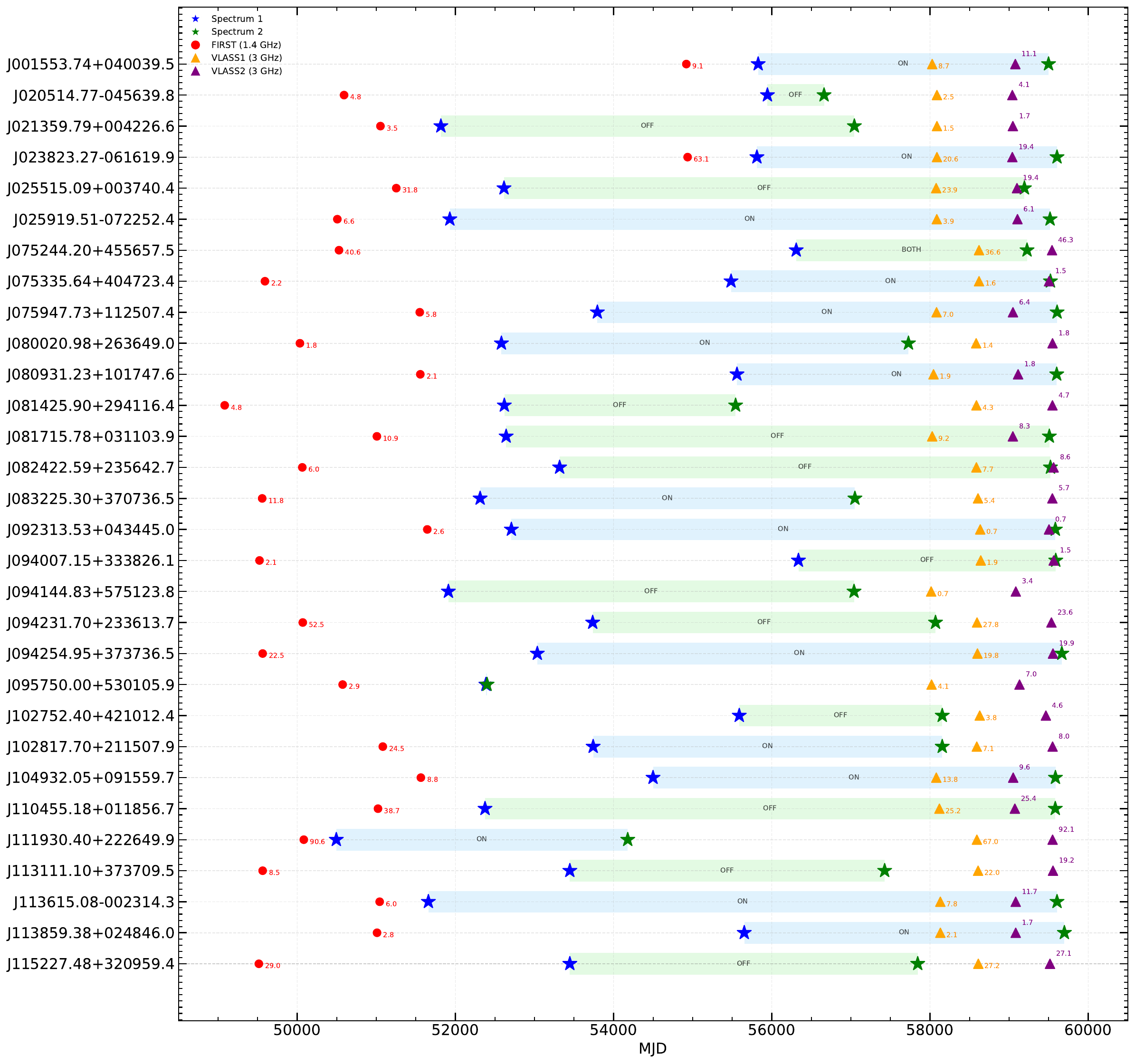}
    \caption{
Radio observation timeline for the 58 CL-AGNs in our sample. Blue and green stars indicate the epochs of the first and second optical spectroscopic observations, respectively. The shaded region between the two spectroscopic epochs represents the interval during which the CL transition occurred. Blue shaded bars correspond to turn-on CL-AGNs, while green shaded bars correspond to turn-off CL-AGNs. Red circles denote radio measurements from the FIRST survey at 1.4 GHz, whereas orange and purple triangles represent observations from the first and second epochs of the VLASS survey at 3 GHz, respectively. The numbers labeled next to the radio symbols indicate the observed radio flux densities in units of mJy beam$^{-1}$. Sources are ordered alphabetically by their object names.}
    \label{fig:timeline}
\end{figure*}

To facilitate interpretation of the temporal relationship between radio observations and CL transitions, Figures~\ref{fig:timeline} provide observation timelines for all sources. These timelines display the epochs of optical spectroscopic observations together with the available radio measurements from FIRST and VLASS, allowing a visual assessment of the relative timing between the CL events and radio observations discussed in Section~\ref{Sec:radio delay}.

\begin{deluxetable*}{lcccccccccccc}
\tabletypesize{\scriptsize}
\setlength{\tabcolsep}{3pt}
\renewcommand{\arraystretch}{1.15}
\tablenum{A}
\tablecaption{Radio and Optical Properties of the Radio-detected CL-AGNs
\label{tab:radio_clagn}}

\tablehead{
\colhead{Object Name} &
\colhead{$z$} &
\colhead{$\rm MJD_{spec1}$} &
\colhead{$\rm MJD_{spec2}$} &
\colhead{Spec Tran} &
\colhead{$f_{\rm peak}$} &
\colhead{$f_{\rm int}$} &
\colhead{RMS} &
\colhead{MJD} &
\colhead{Survey} &
\colhead{$\log L_{5100}$} &
\colhead{$\log M_{\rm BH}$}
}

\startdata
\multirow{2}{*}{J001553.74+040039.5} &
\multirow{2}{*}{0.76} &
\multirow{2}{*}{55827} &
\multirow{2}{*}{59498} &
\multirow{2}{*}{on} &
9.07 & 7.98 & 0.13 & 54919 & FIRST & -- & -- \\
& & & & &
$8.72 \pm 0.23$ & $9.57 \pm 0.42$ & -- & 58027 & VLASS1 & $\ldots$ & -- \\
& & & & &
$11.12 \pm 0.21$ & $11.40 \pm 0.37$ & -- & 59078 & VLASS2 & $\ldots$ & $8.66 \pm 0.13$ \\
\hline
\multirow{2}{*}{J020514.77$-$045639.8} &
\multirow{2}{*}{0.36} &
\multirow{2}{*}{56660} &
\multirow{2}{*}{55944} &
\multirow{2}{*}{off} &
4.79 & 6.43 & 0.16 & 50592 & FIRST & -- & -- \\
& & & & &
$2.47 \pm 0.12$ & $2.56 \pm 0.24$ & -- & 58087 & VLASS1 & $43.90 \pm 0.02$ & -- \\
& & & & &
$4.11 \pm 0.17$ & $5.07 \pm 0.35$ & -- & 59040 & VLASS2 & $43.79 \pm 0.02$ & $8.19 \pm 0.05$ \\
\hline
\multirow{2}{*}{J021359.79+004226.6} &
\multirow{2}{*}{0.18} &
\multirow{2}{*}{57043} &
\multirow{2}{*}{51816} &
\multirow{2}{*}{off} &
3.46 & 3.05 & 0.11 & 51052 & FIRST & -- & -- \\
& & & & &
$1.47 \pm 0.16$ & $3.31 \pm 0.49$ & -- & 58087 & VLASS1 & $43.88 \pm 0.01$ & -- \\
& & & & &
$1.72 \pm 0.15$ & $1.91 \pm 0.29$ & -- & 59047 & VLASS2 & $43.86 \pm 0.01$ & $8.35 \pm 0.01$ \\
\hline
\multirow{2}{*}{J023823.27$-$061619.9} &
\multirow{2}{*}{0.48} &
\multirow{2}{*}{55811} &
\multirow{2}{*}{59605} &
\multirow{2}{*}{on} &
63.08 & 80.75 & 0.13 & 54935 & FIRST & -- & -- \\
& & & & &
$20.64 \pm 0.16$ & $98.57 \pm 1.15$ & -- & 58087 & VLASS1 & $44.15 \pm 0.02$ & -- \\
& & & & &
$19.41 \pm 0.21$ & $97.25 \pm 1.67$ & -- & 59040 & VLASS2 & $44.33 \pm 0.01$ & $8.65 \pm 0.16$ \\
\hline
\multirow{2}{*}{J025515.09+003740.4} &
\multirow{2}{*}{1.01} &
\multirow{2}{*}{59193} &
\multirow{2}{*}{52614} &
\multirow{2}{*}{off} &
31.77 & 32.13 & 0.10 & 51252 & FIRST & -- & -- \\
& & & & &
$23.91 \pm 0.09$ & $24.05 \pm 0.16$ & -- & 58077 & VLASS1 & $\ldots$ & -- \\
& & & & &
$19.41 \pm 0.16$ & $20.43 \pm 0.29$ & -- & 59099 & VLASS2 & $\ldots$ & $8.04 \pm 0.43$ \\
\hline
\multirow{2}{*}{J025919.51-072252.4} &
\multirow{2}{*}{0.54} &
\multirow{2}{*}{51929} &
\multirow{2}{*}{59517} &
\multirow{2}{*}{on} &
6.62 & 7.11 & 0.15 & 50507 & FIRST & -- & -- \\
& & & & &
$3.94 \pm 0.14$ & $4.72 \pm 0.28$ & -- & 58085 & VLASS1 & $44.02 \pm 0.03$ & -- \\
& & & & &
$6.14 \pm 0.15$ & $7.85 \pm 0.38$ & -- & 59105 & VLASS2 & $44.12 \pm 0.02$ & $8.08 \pm 0.13$ \\
\hline
\multirow{2}{*}{J075244.20+455657.5} &
\multirow{2}{*}{0.05} &
\multirow{2}{*}{59226} &
\multirow{2}{*}{56308} &
\multirow{2}{*}{on} &
40.56 & 49.13 & 0.13 & 50528 & FIRST & -- & -- \\
& & & & &
$36.59 \pm 0.22$ & $39.39 \pm 0.62$ & -- & 58619 & VLASS1 & 43.12 & -- \\
& & & & &
$46.35 \pm 0.16$ & $49.32 \pm 0.28$ & -- & 59542 & VLASS2 & 43.08 & $9.11 \pm 0.12$ \\
\hline
\multirow{2}{*}{J075335.64+404723.4} &
\multirow{2}{*}{0.66} &
\multirow{2}{*}{55484} &
\multirow{2}{*}{59522} &
\multirow{2}{*}{on} &
2.16 & 2.74 & 0.14 & 49592 & FIRST & -- & -- \\
& & & & &
$1.62 \pm 0.13$ & $1.66 \pm 0.24$ & -- & 58619 & VLASS1 & $44.05 \pm 0.03$ & -- \\
& & & & &
$1.55 \pm 0.12$ & $1.81 \pm 0.25$ & -- & 59504 & VLASS2 & $44.14 \pm 0.03$ & $8.78 \pm 0.76$ \\
\hline
\multirow{2}{*}{J075947.73+112507.4} &
\multirow{2}{*}{0.34} &
\multirow{2}{*}{53794} &
\multirow{2}{*}{59606} &
\multirow{2}{*}{on} &
5.84 & 5.11 & 0.14 & 51549 & FIRST & -- & -- \\
& & & & &
$7.01 \pm 0.16$ & $7.67 \pm 0.31$ & -- & 58082 & VLASS1 & $43.77 \pm 0.02$ & -- \\
& & & & &
$6.36 \pm 0.21$ & $5.94 \pm 0.36$ & -- & 59048 & VLASS2 & $43.88 \pm 0.01$ & $8.41 \pm 0.01$ \\
\hline
\multirow{2}{*}{J080020.98+263649.0} &
\multirow{2}{*}{0.03} &
\multirow{2}{*}{57727} &
\multirow{2}{*}{52581} &
\multirow{2}{*}{on} &
1.81 & 3.13 & 0.14 & 50034 & FIRST & -- & -- \\
& & & & &
$1.40 \pm 0.13$ & $2.56 \pm 0.35$ & -- & 58583 & VLASS1 & $42.17 \pm 0.01$ & -- \\
& & & & &
$1.80 \pm 0.12$ & $2.74 \pm 0.28$ & -- & 59548 & VLASS2 & $42.11 \pm 0.01$ & $8.00 \pm 0.28$ \\
\hline
\multirow{3}{*}{J080931.23+101747.6} &
\multirow{3}{*}{0.48} &
\multirow{3}{*}{55559} &
\multirow{3}{*}{59601} &
\multirow{3}{*}{on} &
2.1 & 4.94 & 0.15 & 51556 & FIRST & -- & -- \\
& & & & &
$1.86 \pm 0.11$ & $1.50 \pm 0.17$ & -- & 58044 & VLASS1 & $43.92 \pm 0.03$ & -- \\
& & & & &
$2.12 \pm 0.3$ & -- & -- & 59113 & VLASS2 & $44.05 \pm 0.02$ & $8.32 \pm 0.22$ \\
\hline
\multirow{3}{*}{J081425.90+294116.4} &
\multirow{3}{*}{0.37} &
\multirow{3}{*}{55542} &
\multirow{3}{*}{52618} &
\multirow{3}{*}{off} &
4.83 & 4.69 & 0.14 & 49083 & FIRST & -- & -- \\
& & & & &
$4.26 \pm 0.2$ & $4.37 \pm 0.35$ & -- & 58586 & VLASS1 & $44.02 \pm 0.01$ & -- \\
& & & & &
$4.80 \pm 0.21$ & -- & -- & 59547 & VLASS2 & $44.05 \pm 0.01$ & $8.87 \pm 0.05$ \\
\hline
\multirow{3}{*}{J081715.78+031103.9} &
\multirow{3}{*}{0.41} &
\multirow{3}{*}{52641} &
\multirow{3}{*}{59508} &
\multirow{3}{*}{off} &
10.88 & 10.93 & 0.15 & 51007 & FIRST & -- & -- \\
& & & & &
$9.20 \pm 0.14$ & $10.17 \pm 0.26$ & -- & 58027 & VLASS1 & $43.78 \pm 0.03$ & -- \\
& & & & &
$8.56 \pm 0.29$ & -- & -- & 59047 & VLASS2 & $43.84 \pm 0.02$ & $8.71 \pm 0.14$ \\
\hline
\multirow{3}{*}{J082422.59+235642.7} &
\multirow{3}{*}{0.34} &
\multirow{3}{*}{53317} &
\multirow{3}{*}{59522} &
\multirow{3}{*}{off} &
5.95 & 5.28 & 0.16 & 50064 & FIRST & -- & -- \\
& & & & &
$7.71 \pm 0.2$ & $7.85 \pm 0.35$ & -- & 58586 & VLASS1 & $43.91 \pm 0.01$ & -- \\
& & & & &
$8.75 \pm 0.33$ & -- & -- & 59562 & VLASS2 & $43.72 \pm 0.02$ & $8.42 \pm 0.14$ \\
\hline
\multirow{3}{*}{J083225.30+370736.5} &
\multirow{3}{*}{0.09} &
\multirow{3}{*}{57050} &
\multirow{3}{*}{52312} &
\multirow{3}{*}{on} &
11.77 & 11.73 & 0.18 & 49557 & FIRST & -- & -- \\
& & & & &
$5.44 \pm 0.13$ & $5.92 \pm 0.24$ & -- & 58606 & VLASS1 & $43.94$ & -- \\
& & & & &
$5.57 \pm 0.20$ & -- & -- & 59546 & VLASS2 & $43.91$ & $9.41 \pm 0.01$ \\
\hline
\multirow{3}{*}{J094007.15+333826.1} &
\multirow{3}{*}{2.56} &
\multirow{3}{*}{56336} &
\multirow{3}{*}{59589} &
\multirow{3}{*}{off} &
2.06 & 2.17 & 0.15 & 49523 & FIRST & -- & -- \\
& & & & &
$1.94 \pm 0.14$ & $2.50 \pm 0.30$ & -- & 58641 & VLASS1 & $45.07 \pm 0.02$ & -- \\
& & & & &
$1.93 \pm 0.25$ & -- & -- & 59563 & VLASS2 & $45.07 \pm 0.04$ & $8.84 \pm 0.21$ \\
\hline
\enddata
\tablecomments{
Each source is listed in two rows. The FIRST and VLASS1 measurements are given first, followed by the VLASS2 measurements and the black hole mass.
}
\end{deluxetable*}

\begin{deluxetable*}{lcccccccccccc}
\tabletypesize{\scriptsize}
\setlength{\tabcolsep}{3pt}
\renewcommand{\arraystretch}{1.15}
\tablenum{A}
\tablecaption{Continued}

\tablehead{
\colhead{Object Name} &
\colhead{$z$} &
\colhead{$\rm MJD_{spec1}$} &
\colhead{$\rm MJD_{spec2}$} &
\colhead{Spec Tran} &
\colhead{$f_{\rm peak}$} &
\colhead{$f_{\rm int}$} &
\colhead{RMS} &
\colhead{MJD} &
\colhead{Survey} &
\colhead{$\log L_{5100}$} &
\colhead{$\log M_{\rm BH}$}
}

\startdata
\multirow{3}{*}{J094231.70+233613.7} &
\multirow{3}{*}{0.80} &
\multirow{3}{*}{58069} &
\multirow{3}{*}{53735} &
\multirow{3}{*}{off} &
52.49 & 152.98 & 0.15 & 50070 & FIRST & -- & -- \\
& & & & &
$27.84 \pm 0.18$ & $44.24 \pm 0.73$ & -- & 58594 & VLASS1 & $45.11 \pm 0.01$ & -- \\
& & & & &
$68.34 \pm 1.07$ & -- & -- & 59534 & VLASS2 & $45.11 \pm 0.01$ & 9.60 \\
\hline
\multirow{3}{*}{J094254.95+373736.5} &
\multirow{3}{*}{0.50} &
\multirow{3}{*}{53035} &
\multirow{3}{*}{59666} &
\multirow{3}{*}{on} &
22.46 & 43.32 & 0.15 & 49563 & FIRST & -- & -- \\
& & & & &
$19.85 \pm 0.13$ & $22.98 \pm 0.36$ & -- & 58599 & VLASS1 & $43.97 \pm 0.03$ & -- \\
& & & & &
$22.46 \pm 0.34$ & -- & -- & 59554 & VLASS2 & $43.85 \pm 0.04$ & 8.468 \\
\hline
\multirow{3}{*}{J095750.00+530105.9} &
\multirow{3}{*}{0.44} &
\multirow{3}{*}{52400} &
\multirow{3}{*}{52385} &
\multirow{3}{*}{on} &
2.92 & 2.50 & 0.14 & 50574 & FIRST & -- & -- \\
& & & & &
$4.11 \pm 0.12$ & $4.49 \pm 0.23$ & -- & 58019 & VLASS1 & $44.71 \pm 0.01$ & -- \\
& & & & &
$7.31 \pm 0.24$ & -- & -- & 59131 & VLASS2 & 44.68 & $8.68 \pm 0.02$ \\
\hline
\multirow{3}{*}{J102817.70+211507.9} &
\multirow{3}{*}{0.36} &
\multirow{3}{*}{58154} &
\multirow{3}{*}{53741} &
\multirow{3}{*}{on} &
24.47 & 30.93 & 0.14 & 51082 & FIRST & -- & -- \\
& & & & &
$7.08 \pm 0.24$ & $6.01 \pm 0.37$ & -- & 58591 & VLASS1 & $44.31 \pm 0.01$ & -- \\
& & & & &
$9.56 \pm 0.31$ & -- & -- & 59549 & VLASS2 & $44.47 \pm 0.01$ & $8.01 \pm 0.02$ \\
\hline
\multirow{3}{*}{J104932.05+091559.7} &
\multirow{3}{*}{0.51} &
\multirow{3}{*}{54498} &
\multirow{3}{*}{59585} &
\multirow{3}{*}{on} &
8.76 & 14.82 & 0.14 & 51563 & FIRST & -- & -- \\
& & & & &
$13.84 \pm 0.12$ & $13.05 \pm 0.21$ & -- & 58079 & VLASS1 & $44.71 \pm 0.01$ & -- \\
& & & & &
$9.04 \pm 0.26$ & -- & -- & 59051 & VLASS2 & $44.65 \pm 0.01$ & $8.85 \pm 0.21$ \\
\hline
\multirow{3}{*}{J110455.18+011856.7} &
\multirow{3}{*}{0.57} &
\multirow{3}{*}{52374} &
\multirow{3}{*}{59583} &
\multirow{3}{*}{off} &
38.74 & 39.72 & 0.15 & 51021 & FIRST & -- & -- \\
& & & & &
$25.22 \pm 0.13$ & $25.99 \pm 0.23$ & -- & 58118 & VLASS1 & $43.47 \pm 0.13$ & -- \\
& & & & &
$25.28 \pm 0.41$ & -- & -- & 59072 & VLASS2 & $43.79 \pm 0.05$ & $8.78 \pm 0.13$ \\
\hline
\multirow{3}{*}{J111930.40+222649.9} &
\multirow{3}{*}{0.42} &
\multirow{3}{*}{54178} &
\multirow{3}{*}{50494} &
\multirow{3}{*}{on} &
90.58 & 92.75 & 0.15 & 50083 & FIRST & -- & -- \\
& & & & &
$66.98 \pm 0.13$ & $69.77 \pm 0.29$ & -- & 58592 & VLASS1 & 44.89 & -- \\
& & & & &
$96.027 \pm 0.22$ & -- & -- & 59549 & VLASS2 & 45.05 & 9.20 \\
\hline
\multirow{3}{*}{J113111.10+373709.5} &
\multirow{3}{*}{0.45} &
\multirow{3}{*}{57426} &
\multirow{3}{*}{53446} &
\multirow{3}{*}{off} &
8.49 & 9.77 & 0.15 & 49563 & FIRST & -- & -- \\
& & & & &
$21.96 \pm 0.13$ & $22.17 \pm 0.24$ & -- & 58607 & VLASS1 & $44.40 \pm 0.01$ & -- \\
& & & & &
$19.96 \pm 0.22$ & -- & -- & 59554 & VLASS2 & 44.36 & $9.35 \pm 0.24$ \\
\hline
\multirow{3}{*}{J113615.08-002314.3} &
\multirow{3}{*}{0.39} &
\multirow{3}{*}{51658} &
\multirow{3}{*}{59605} &
\multirow{3}{*}{on} &
5.97 & 6.34 & 0.17 & 51042 & FIRST & -- & -- \\
& & & & &
$7.79 \pm 0.16$ & $7.58 \pm 0.28$ & -- & 58130 & VLASS1 & $44.32 \pm 0.01$ & -- \\
& & & & &
$10.63 \pm 0.31$ & -- & -- & 59083 & VLASS2 & $44.33 \pm 0.01$ & $8.99 \pm 0.18$ \\
\hline
\multirow{3}{*}{J113859.38+024846.0} &
\multirow{3}{*}{0.45} &
\multirow{3}{*}{55651} &
\multirow{3}{*}{59698} &
\multirow{3}{*}{on} &
2.78 & 3.75 & 0.16 & 51009 & FIRST & -- & -- \\
& & & & & $2.13 \pm 0.16$ & $2.98 \pm 0.34$ & -- & 58130 & VLASS1 & $43.63 \pm 0.05$ & --  \\
& & & & & $1.68 \pm 0.19$ &
$6.42 \pm 0.89$ & -- & 59083 & VLASS2 & $43.85 \pm 0.02$ & $8.53 \pm 0.16$ \\
\hline
\multirow{3}{*}{J115227.48+320959.4} &
\multirow{3}{*}{0.37} &
\multirow{3}{*}{57844} &
\multirow{3}{*}{53446} &
\multirow{3}{*}{off} &
29.02 & 32.92 & 0.15 & 49515 & FIRST & -- & -- \\
& & & & & $27.16 \pm 0.13$ & $32.18 \pm 0.42$ & -- & 58610 & VLASS1 & 44.35 & -- \\
& & & & & $27.12 \pm 0.12$ & 
$30.62 \pm 0.26$ & -- & 59516 & VLASS2 & $44.34 \pm 0.01$ & $8.51 \pm 0.14$ \\
\hline
\multirow{3}{*}{J120349.21+605317.4} &
\multirow{3}{*}{0.07} &
\multirow{3}{*}{59696} &
\multirow{3}{*}{52405} &
\multirow{3}{*}{on} &
1.07 & 0.95 & 0.15 & 52467 & FIRST & -- & -- \\
& & & & &
$1.20 \pm 0.15$ & $2.06 \pm 0.39$ & -- & 58014 & VLASS1 & $42.81 \pm 0.01$ & --\\
& & & & & $1.24 \pm 0.13$ &
$1.72 \pm 0.29$ &  -- & 59026 & VLASS2 & $42.67 \pm 0.01$ & $\ldots$ \\
\hline
\multirow{3}{*}{J120356.94+345941.6} &
\multirow{3}{*}{0.56} &
\multirow{3}{*}{55621} &
\multirow{3}{*}{59587} &
\multirow{3}{*}{on} &
29.64 & 31.59 & 0.14 & 49536 & FIRST &--&-- \\
& & & & &
$18.77 \pm 0.14$ & $20.69 \pm 0.26$ & -- & 58089 & VLASS1 & $44.15 \pm 0.01$ & --\\
& & & & & $19.47 \pm 0.16$ &
$19.87 \pm 0.30$ & --  & 59104 & VLASS2 & $44.22 \pm 0.01$ & $8.09 \pm 0.25$ \\
\hline
\multirow{3}{*}{J120435.12+485532.9} &
\multirow{3}{*}{0.80} &
\multirow{3}{*}{57135} &
\multirow{3}{*}{59657} &
\multirow{3}{*}{on} &
12.72 & 12.52 & 0.15 & 50552 & FIRST &--&-- \\
& & & & &
$13.66 \pm 0.13$ & $14.49 \pm 0.31$ & -- & 58077 & VLASS1 & $44.47 \pm 0.01$ & --\\
& & & & &$13.80 \pm 0.26$ &
$15.96 \pm 0.49$  & -- & 59062 & VLASS2 & $44.60 \pm 0.01$ & $8.74 \pm 0.30$ \\
\hline
\multirow{3}{*}{J122550.31+510846.3} &
\multirow{3}{*}{0.17} &
\multirow{3}{*}{58423} &
\multirow{3}{*}{$\ldots$} &
\multirow{3}{*}{on} &
2.17 & 2.59 & 0.14 & 50565 & FIRST &--&-- \\
& & & & &
$2.59 \pm 0.12$ & $2.48 \pm 0.21$ & -- & 58077 & VLASS1 & 43.57 & -- \\
& & & & &$2.86 \pm 0.16$&
$3.42 \pm 0.32$  & -- & 59062 & VLASS2 & $43.63 \pm 0.01$ & $9.11 \pm 0.10$ \\
\hline
\multirow{3}{*}{J123819.60+412420.5} &
\multirow{3}{*}{0.50} &
\multirow{3}{*}{57511} &
\multirow{3}{*}{53090} &
\multirow{3}{*}{off} &
8.08 & 8.20 & 0.15 & 49599 & FIRST &--&--  \\
& & & & &
$10.58 \pm 0.13$ & $10.91 \pm 0.23$ & -- & 58588 & VLASS1 & 44.26 & -- \\
& & & & &$10.18 \pm 0.13$&
$10.11 \pm 0.22$  & -- & 59532 & VLASS2 & $44.33 \pm 0.01$ & $9.11 \pm 0.13$ \\
\hline
\enddata

\end{deluxetable*}

\begin{deluxetable*}{lcccccccccccc}
\tabletypesize{\scriptsize}
\setlength{\tabcolsep}{3pt}
\renewcommand{\arraystretch}{1.15}
\tablenum{A}
\tablecaption{Continued}

\tablehead{
\colhead{Object Name} &
\colhead{$z$} &
\colhead{$\rm MJD_{spec1}$} &
\colhead{$\rm MJD_{spec2}$} &
\colhead{Spec Tran} &
\colhead{$f_{\rm peak}$} &
\colhead{$f_{\rm int}$} &
\colhead{RMS} &
\colhead{MJD} &
\colhead{Survey} &
\colhead{$\log L_{5100}$} &
\colhead{$\log M_{\rm BH}$}
}

\startdata
\multirow{3}{*}{J132045.25-002449.6} & \multirow{3}{*}{0.22} & \multirow{3}{*}{51578} & \multirow{3}{*}{59653} & \multirow{3}{*}{off} & 3.88 & 4.84 & 0.14 & 51037 & FIRST & -- & -- \\
& & & & &  $6.30 \pm 0.25$ & $7.08 \pm 0.48$ & -- & 58566 & VLASS1 & $43.47 \pm 0.02$ & -- \\
& & & & & $6.08 \pm 0.20$ & $6.67 \pm 0.37$ & -- & 59490 & VLASS2 & 43.37 & $8.19 \pm 0.16$ \\
\hline
\multirow{3}{*}{J134133.70+090356.2} &
\multirow{3}{*}{0.10} &
\multirow{3}{*}{58154} &
\multirow{3}{*}{53886} &
\multirow{3}{*}{off} &
19.7 & 21.86 & 0.21 & 51565 & FIRST &--&-- \\
& & & & &
$7.99 \pm 0.15$ & $9.34 \pm 0.29$ & -- & 58595 & VLASS1 & 43.51 &-- \\
& & & & & $7.18 \pm 0.13$&
$9.12 \pm 0.27$  & -- & 59552 & VLASS2 & 43.49 & $7.83 \pm 0.07$ \\
\hline
\multirow{3}{*}{J134628.40+192243.3} &
\multirow{3}{*}{0.08} &
\multirow{3}{*}{58662} &
\multirow{3}{*}{54507} &
\multirow{3}{*}{off} &
1.11 & 0.90 & 0.14 & 51498 & FIRST &--&--  \\
& & & & &
$1.13 \pm 0.13$ & $1.42 \pm 0.28$ & -- & 58574 & VLASS1 & 43.62 & --\\
& & & & &$0.92 \pm 0.13$ &
$2.28 \pm 0.44$  & -- & 59502 & VLASS2 & 43.59 & 8.50 \\
\hline
\multirow{3}{*}{J135359.47+084526.7} &
\multirow{3}{*}{0.15} &
\multirow{3}{*}{53559} &
\multirow{3}{*}{59709} &
\multirow{3}{*}{on} &
3.07 & 2.51 & 0.14 & 51567 & FIRST & --&-- \\
& & & & &
$6.80 \pm 0.14$ & $6.80 \pm 0.24$ & -- & 58595 & VLASS1 & $43.00 \pm 0.02$ & -- \\
& & & & &$7.27 \pm 0.12$&
$7.50 \pm 0.22$  & -- & 59552 & VLASS2 & $42.96 \pm 0.02$ & $8.27 \pm 0.13$ \\
\hline
\multirow{3}{*}{J140951.40-031633.7} &
\multirow{3}{*}{0.82} &
\multirow{3}{*}{55383} &
\multirow{3}{*}{59623} &
\multirow{3}{*}{on} &
2.86 & 2.40 & 0.15 & 51069 & FIRST & --&-- \\
& & & & &
$3.24 \pm 0.16$ & $3.51 \pm 0.29$ & -- & 58604 & VLASS1 & $44.49 \pm 0.02$ & -- \\
& & & & &$3.81 \pm 0.15$&
$4.26 \pm 0.29$  & -- & 59511 & VLASS2 & $44.51 \pm 0.02$ & 9.06 \\
\hline
\multirow{3}{*}{J141403.20+352311.8} &
\multirow{3}{*}{0.06} &
\multirow{3}{*}{58694} &
\multirow{3}{*}{53143} &
\multirow{3}{*}{on} &
24.37 & 25.01 & 0.14 & 49538 & FIRST &--&--  \\
& & & & &
$11.96 \pm 0.11$ & $12.89 \pm 0.20$ & -- & 58097 & VLASS1 & 43.30 & -- \\
& & & & &$11.75 \pm 0.15$&
$13.44 \pm 0.29$  & -- & 59107 & VLASS2 & 43.36 & $7.00 \pm 0.01$ \\
\hline
\multirow{3}{*}{J142852.80+271043.0} &
\multirow{3}{*}{0.44} &
\multirow{3}{*}{57844} &
\multirow{3}{*}{56067} &
\multirow{3}{*}{on} &
6.84 & 6.7362 & 0.17 & 50029 & FIRST &--&--  \\
& & & & &
$6.66 \pm 0.13$ & $7.06 \pm 0.23$ & -- & 58069 & VLASS1 & 44.84 & -- \\
& & & & &$7.63 \pm 0.13$&
$7.58 \pm 0.24$  & -- & 59099 & VLASS2 & 44.70 & 7.70 \\
\hline
\multirow{3}{*}{J145359.74+091543.2} &
\multirow{3}{*}{0.28} &
\multirow{3}{*}{58897} &
\multirow{3}{*}{53827} &
\multirow{3}{*}{off} &
118.06 & 122.0084 & 0.14 & 51564 & FIRST &--&--  \\
& & & & &
$95.18 \pm 0.17$ & $120.84 \pm 0.77$ & -- & 58554 & VLASS1 & $43.97 \pm 0.01$ & --\\
& & & & &$68.66 \pm 0.14$&
$70.82 \pm 0.25$  & -- & 59547 & VLASS2 & $43.91 \pm 0.01$ & $8.03 \pm 0.09$ \\
\hline
\multirow{3}{*}{J145916.84+341433.2} &
\multirow{3}{*}{0.43} &
\multirow{3}{*}{55653} &
\multirow{3}{*}{59629} &
\multirow{3}{*}{on} &
16.07 & 16.85 & 0.14 & 49529 & FIRST &--&--  \\
& & & & &$9.86 \pm 0.14$ & $10.17 \pm 0.26$ & -- & 58084 & VLASS1 & $44.15 \pm 0.01$ & -- \\
& & & & &$12.34 \pm 0.15$&
$12.01 \pm 0.26$  & -- & 59104 & VLASS2 & $44.18 \pm 0.01$ & $8.09 \pm 0.18$ \\
\hline
\multirow{3}{*}{J152641.90+163246.3} &
\multirow{3}{*}{0.83} &
\multirow{3}{*}{58662} &
\multirow{3}{*}{54266} &
\multirow{3}{*}{on} &
93.34 & 95.2752 & 0.14 & 51489 & FIRST &--&--  \\
& & & & &
$108.51 \pm 0.17$ & $112.57 \pm 0.42$ & -- & 58563 & VLASS1 & 45.29 & -- \\
& & & & &$105.44 \pm 0.16$&
$107.61 \pm 0.49$  & -- & 59558 & VLASS2 & 45.32 & 8.60 \\
\hline
\multirow{3}{*}{J153310.02+272920.2} &
\multirow{3}{*}{0.07} &
\multirow{3}{*}{54180} &
\multirow{3}{*}{59684} &
\multirow{3}{*}{on} &
3.13 & 3.043 & 0.13 & 50028 & FIRST &--&--  \\
& & & & &
$1.63 \pm 0.12$ & $1.83 \pm 0.23$ & -- & 58028 & VLASS1 & 43.29 & -- \\
& & & & &$1.81 \pm 0.13$&
$1.68 \pm 0.22$  & -- & 59098 & VLASS2 & 43.09 & $7.73 \pm 0.14$ \\
\hline
\multirow{3}{*}{J153329.95+074102.1} &
\multirow{3}{*}{0.88} &
\multirow{3}{*}{55735} &
\multirow{3}{*}{59732} &
\multirow{3}{*}{on} &
3.75 & 3.49 & 0.15 & 51574 & FIRST &--&--  \\
& & & & &
$2.31 \pm 0.15$ & $2.73 \pm 0.30$ & -- & 58556 & VLASS1 & $44.63 \pm 0.02$ & --\\
& & & & &$2.19 \pm 0.14$ &
$2.52 \pm 0.26$  & -- & 59524 & VLASS2 & $44.70 \pm 0.01$ & $8.75 \pm 0.22$ \\
\hline
\multirow{3}{*}{J153354.60+345504.4} &
\multirow{3}{*}{0.75} &
\multirow{3}{*}{57570} &
\multirow{3}{*}{53144} &
\multirow{3}{*}{off} &
2.38 & 1.84 & 0.14 & 49536 & FIRST &--&--  \\
& & & & &
$2.58 \pm 0.12$ & $2.29 \pm 0.19$ & -- & 58089 & VLASS1 & $44.51 \pm 0.01$ & -- \\
& & & & &$2.43 \pm 0.14$&
$2.71 \pm 0.26$ & -- & 59110 & VLASS2 & $44.61 \pm 0.01$ & $9.32 \pm 0.25$ \\
\hline
\multirow{3}{*}{J153502.47+102150.4} &
\multirow{3}{*}{0.57} &
\multirow{3}{*}{56033} &
\multirow{3}{*}{59393} &
\multirow{3}{*}{on} &
7.40 & 8.31 & 0.14 & 51559 & FIRST &--&--  \\
& & & & &$6.81 \pm 0.13$ & $7.53 \pm 0.25$ & -- & 58556 & VLASS1 & $43.75 \pm 0.05$ & -- \\
& & & & &$8.44 \pm 0.12$&
$8.98 \pm 0.23$ & -- & 59552 & VLASS2 & $43.82 \pm 0.03$ & $8.14 \pm 0.13$ \\
\hline
\multirow{3}{*}{J160002.84+322644.6} &
\multirow{3}{*}{0.32} &
\multirow{3}{*}{55719} &
\multirow{3}{*}{59358} &
\multirow{3}{*}{on} &
5.68 & 7.58 & 0.15 & 49516 & FIRST &--&--  \\
& & & & &
$5.28 \pm 0.19$ & $5.44 \pm 0.33$ & -- & 58031 & VLASS1 & $42.67 \pm 0.14$ & -- \\
& & & & &$6.44 \pm 0.21$&
$6.76 \pm 0.39$  & -- & 59107 & VLASS2 & $43.21 \pm 0.06$ & $8.77 \pm 0.17$ \\
\hline
\multirow{3}{*}{J161413.20+260416.0} &
\multirow{3}{*}{0.13} &
\multirow{3}{*}{52824} &
\multirow{3}{*}{48004} &
\multirow{3}{*}{on} &
17.69 & 17.75 & 0.14 & 50041 & FIRST &--&--  \\
& & & & &
$7.28 \pm 0.12$ & $8.72 \pm 0.24$ & -- & 58081 & VLASS1 & 44.66 & -- \\
& & & & &$7.46 \pm 0.14$&
$8.00 \pm 0.25$  & -- & 59096 & VLASS2 & 44.67 & $8.34 \pm 0.01$ \\
\hline
\enddata

\end{deluxetable*}

\begin{deluxetable*}{lcccccccccccc}
\tabletypesize{\scriptsize}
\setlength{\tabcolsep}{3pt}
\renewcommand{\arraystretch}{1.15}
\tablenum{A}
\tablecaption{Continued}

\tablehead{
\colhead{Object Name} &
\colhead{$z$} &
\colhead{$\rm MJD_{spec1}$} &
\colhead{$\rm MJD_{spec2}$} &
\colhead{Spec Tran} &
\colhead{$f_{\rm peak}$} &
\colhead{$f_{\rm int}$} &
\colhead{RMS} &
\colhead{MJD} &
\colhead{Survey} &
\colhead{$\log L_{5100}$} &
\colhead{$\log M_{\rm BH}$}
}

\startdata
\multirow{3}{*}{J162552.80+125316.4} &
\multirow{3}{*}{0.38} &
\multirow{3}{*}{58662} &
\multirow{3}{*}{53881} &
\multirow{3}{*}{on} &
2.11 & 1.37 & 0.14 & 51545 & FIRST &--&--  \\
& & & & &
$2.68 \pm 0.13$ & $2.40 \pm 0.22$ & -- & 58588 & VLASS1 & 44.44 & -- \\
& & & & &$2.26 \pm 0.13$&
$2.27 \pm 0.22$  & -- & 59490 & VLASS2 & 44.50 & 9.30 \\
\hline
\multirow{3}{*}{J163959.17+511930.9} &
\multirow{3}{*}{1.13} &
\multirow{3}{*}{57190} &
\multirow{3}{*}{59383} &
\multirow{3}{*}{on} &
5.32 & 5.38 & 0.13 & 50566 & FIRST &--&--  \\
& & & & &
$5.16 \pm 0.12$ & $5.21 \pm 0.20$ & -- & 58042 & VLASS1 & $\ldots$ & -- \\
& & & & &$6.06 \pm 0.15$&
$6.13 \pm 0.26$  & -- & 59072 & VLASS2 & $\ldots$ & $9.20 \pm 0.02$ \\
\hline
\multirow{3}{*}{J164331.91+304835.6} &
\multirow{3}{*}{0.18} &
\multirow{3}{*}{52793} &
\multirow{3}{*}{59386} &
\multirow{3}{*}{off} &
63.76 & 65.48 & 0.15 & 49533 & FIRST &--&--  \\
& & & & &
$45.47 \pm 0.12$ & $47.84 \pm 0.21$ & -- & 58082 & VLASS1 & $43.30 \pm 0.01$ & -- \\
& & & & &$45.64 \pm 0.15$&
$48.19 \pm 0.26$  & -- & 59103 & VLASS2 & $43.26 \pm 0.01$ & $7.36 \pm 0.13$ \\
\hline
\multirow{3}{*}{J171602.00+311214.0} &
\multirow{3}{*}{0.11} &
\multirow{3}{*}{52431} &
\multirow{3}{*}{50551} &
\multirow{3}{*}{on} &
2.68 & 2.42 & 0.13 & 49656 & FIRST &--&--  \\
& & & & &
1.37 & 1.37 & -- & 58027 & VLASS1 & 44.39 & -- \\
& & & & &$1.51 \pm 0.14$&
$1.66 \pm 0.27$  & -- & 59103 & VLASS2 & 44.41 & 7.80 \\
\hline
\multirow{3}{*}{J214613.31+000931.0} &
\multirow{3}{*}{0.62} &
\multirow{3}{*}{55478} &
\multirow{3}{*}{52968} &
\multirow{3}{*}{on} &
8.88 & 9.02 & 0.10 & 51248 & FIRST &--&--  \\
& & & & &
$6.96 \pm 0.10$ & $7.44 \pm 0.18$ & -- & 58023 & VLASS1 & $\ldots$ & -- \\
& & & & &$5.53 \pm 0.15$&
$6.47 \pm 0.30$  & -- & 59049 & VLASS2 & $44.26 \pm 0.01$ & $9.17 \pm 0.08$ \\
\hline
\multirow{3}{*}{J231733.18-100504.3} &
\multirow{3}{*}{0.59} &
\multirow{3}{*}{52203} &
\multirow{3}{*}{59516} &
\multirow{3}{*}{off} &
3.07 & 2.42 & 0.14 & 50554 & FIRST &--&--  \\
& & & & &
1.21 & 1.21 & -- & 58647 & VLASS1 & $44.18 \pm 0.02$ & -- \\
& & & & &1.49&
1.49  & -- & 59609 & VLASS2 & $44.16 \pm 0.02$ & $8.64 \pm 0.14$ \\
\hline
\multirow{3}{*}{J092313.53+043445.0} & \multirow{3}{*}{0.66} & \multirow{3}{*}{52707} & \multirow{3}{*}{59584} & \multirow{3}{*}{on} & 2.57 & 3.95 & 0.15 & 51644 & FIRST & $44.34 \pm 0.03$ & -- \\
& & & & & \textless{0.7} & \textless{0.7} & -- & 58634 & VLASS1 & $\ldots$ & -- \\
& & & & & \textless{0.7} & \textless{0.7} & -- & 59504 & VLASS2 & $44.80 \pm 0.01$ & $9.28 \pm 0.18$ \\
\hline
\multirow{3}{*}{J094144.83+575123.8} & \multirow{3}{*}{0.16} & \multirow{3}{*}{51911} & \multirow{3}{*}{57039} & \multirow{3}{*}{off} & $\textless{1.0}$ & $\textless{1.0}$ & $\ldots$ & -- & FIRST & $43.66 \pm 0.07$ & -- \\
& & & & & $\textless{0.7}$ & $\textless{0.7}$ & -- & 58014 & VLASS1 & $\ldots$ & -- \\
& & & & & $3.40 \pm 0.17$ & $4.71 \pm 0.36$ & -- & 59084 & VLASS2 & $43.08 \pm 0.07$ & $8.29 \pm 0.05$ \\
\hline
\multirow{3}{*}{J102752.40+421012.4} & \multirow{3}{*}{0.93} & \multirow{3}{*}{55588} & \multirow{3}{*}{58154} & \multirow{3}{*}{off} & $\textless{1.0}$ & $\textless{1.0}$ & $\ldots$ & -- & FIRST & 44.68 & -- \\
& & & & & $3.78 \pm 0.15$ & $4.03 \pm 0.28$ & -- & 58628 & VLASS1 & $\ldots$ & -- \\
& & & & & $4.61 \pm 0.12$ & $4.81 \pm 0.21$ & -- & 59464 & VLASS2 & $\ldots$ & 8.4 \\
\hline
\multirow{3}{*}{J230443.60-084110.0} & \multirow{3}{*}{0.05} & \multirow{3}{*}{52258} & \multirow{3}{*}{58013} & \multirow{3}{*}{on} & 17.69 & 22.48 & 0.16 & 50500 & FIRST & 42.56 & -- \\
& & & & & $8.40 \pm 0.14$ & $8.88 \pm 0.28$ & -- & 58635 & VLASS1 & $\ldots$ & -- \\
& & & & & $\textless{0.7}$ & $\textless{0.7}$ & -- & 59609 & VLASS2 & 43.59 & 7.6 \\
\hline
\enddata
\end{deluxetable*}

\begin{figure*}
    \centering
    \includegraphics[width=0.98\linewidth]{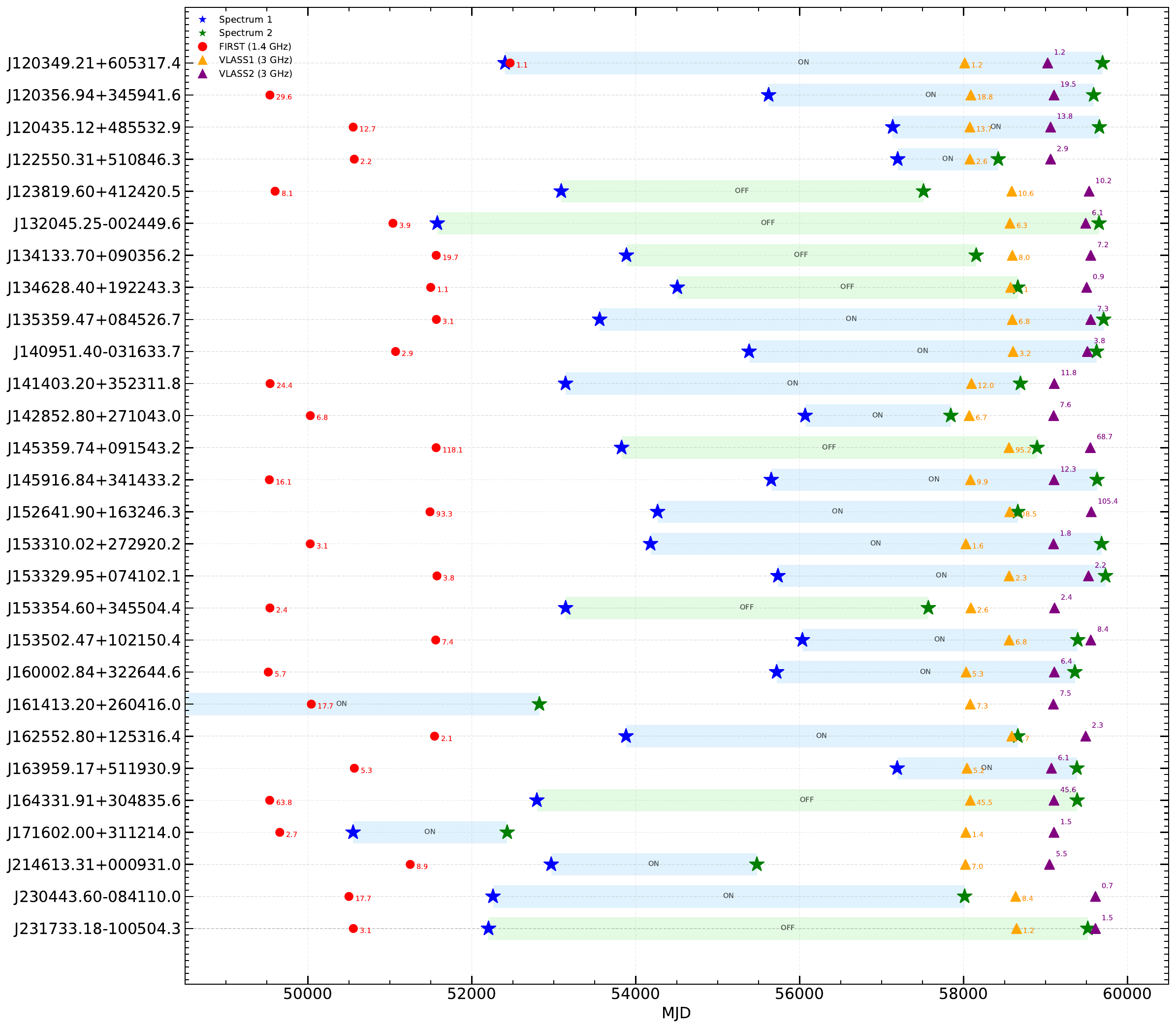}
    \caption{Radio observation timeline for the CL-AGN sample (continued).}
\end{figure*}

\bibliography{Jet_CLAGNs}{}
\bibliographystyle{aasjournalv7}



\end{document}